\documentclass[aoas]{imsart}

\RequirePackage{amsthm,amsmath,amsfonts,amssymb}
\RequirePackage[authoryear]{natbib}
\usepackage[linesnumbered,ruled]{algorithm2e}
\RequirePackage[colorlinks,citecolor=blue,urlcolor=blue]{hyperref}
\RequirePackage{graphicx}
\usepackage{xcolor}

\usepackage{algpseudocode}
\usepackage{cleveref}
\usepackage{booktabs}

\newcommand*{\bs}{\boldsymbol}
\newcommand*{\mb}{\mathbf}

\usepackage{soul} 

\startlocaldefs
\theoremstyle{plain}
\newtheorem{axiom}{Axiom}
\newtheorem{claim}[axiom]{Claim}
\newtheorem{theorem}{Theorem}[section]

\newtheorem{lemma}[theorem]{Lemma}
\theoremstyle{definition}
\newtheorem{definition}[theorem]{Definition}
\newtheorem*{example}{Example}
\newtheorem*{fact}{Fact}

\endlocaldefs

\begin{document}

\begin{frontmatter}
\title{Bayesian  Mixed-Effects Models  for Multilevel Two-way Functional Data: Applications to EEG Experiments}

\begin{aug}
\author[A]{\fnms{Xiaomeng}~\snm{Ju}\ead[label=e1]{jux01@nyu.edu}},
\author[A]{\fnms{Thaddeus}~\snm{Tarpey}\ead[label=e2]{Thaddeus.Tarpey@nyulangone.org}}
\and 
\author[A]{\fnms{Hyung G}~\snm{Park}\ead[label=e3]{parkh15@nyu.edu}} 
\address[A]{Division of Biostatistics, Department of Population Health, New York University \printead[presep={ ,\ }]{e1,e2,e3}}
\end{aug}

\begin{abstract}
In multi-condition EEG experiments, brain activity is recorded  as subjects perform various tasks or are exposed to different stimuli. The recorded signals are commonly transformed into time-frequency representations, which often display smooth variations across time and frequency dimensions.  These representations are naturally structured as two-way functional data, with experimental conditions nested within subjects.  Existing analytical methods fail to jointly account for the data’s multilevel structure, functional nature, and dependence on subject-level covariates. To address these limitations, we propose a Bayesian mixed-effects model for two-way functional data that incorporates covariate-dependent fixed effects at the condition level and multilevel random effects. For enhanced model interpretability and parsimony, we introduce a novel covariate-dependent CANDECOMP/PARAFAC (CP) decomposition for the fixed effects,  with marginally interpretable time and frequency patterns.   We further propose a sparsity-inducing prior for CP rank selection and an efficient algorithm for posterior sampling.  The proposed method is evaluated through extensive simulations and applied to EEG data collected to investigate the effects of alcoholism on cognitive processing in response to visual stimuli.  Our analysis reveals distinct patterns of time-frequency activity associated with alcoholism, offering new insights into the neural processing differences between subject groups and experimental conditions. 
\end{abstract}


\end{frontmatter}

\section{Introduction}

Electroencephalography (EEG) is a non-invasive neuroimaging procedure that records electrical voltage on the scalp with high temporal resolution. Many EEG studies adopt a multi-condition design, where each subject (i.e., participant) performs different tasks or is exposed to various stimuli, each of which defines an experimental condition \citep{malik2017designing}. EEG signals are recorded for each condition per subject, resulting in a multilevel data structure. These signals are commonly transformed into the time-frequency domain to analyze dynamic neural oscillatory patterns \citep{morales2022time}, which typically vary smoothly over the time and frequency domains and can be effectively modeled as two-way functional data.    In many studies, EEG data are supplemented by subject-level covariates, such as demographic variables or clinical diagnoses, which may influence neural responses to experimental conditions. The goal is to uncover how brain responses vary by condition as a function of covariates in the time-frequency domain, while separating condition-specific effects from subject-level variability.  This task presents statistical challenges in multi-condition EEG experiments due to the data’s multilevel structure, two-way functional nature, and dependence on auxiliary covariates.  While existing methods have addressed one or two of these challenges, there lacks a unified solution that tackles all three challenges effectively.

A variety of methods have been developed to analyze multi-condition EEG data, however, many do not fully account for the multilevel structure inherent in such datasets. Some approaches jointly model multiple conditions for a single subject, allowing information sharing across conditions  \citep{gramfort2009improving, wu2014bayesian}, but they are not suited for multi-subject studies where borrowing information across individuals is beneficial. For multi-subject studies, methods based on tensor decomposition \citep{cong2015tensor} and Gaussian processes \citep{marquand2014bayesian} have been explored to capture intra-subject correlations. However, these methods do not explicitly separate population-level condition effects from subject-specific effects reflecting individual variability, and are unable to accommodate auxiliary covariates or functional features.

Mixed-effects models provide an interpretable and flexible framework for EEG analysis, naturally accommodating multilevel data structures in which measurements are nested within subjects across experimental conditions, repeated trials, or brain regions. Most developments of mixed-effects models in EEG studies consider scalar or vector features extracted from raw signals \citep{spinnato2015detecting,riha2020accounting,meinardi2024linear,palma2024understanding}. More recently, mixed-effects models have been extended to functional EEG data to capture continuous changes of signals over time. These extensions include parametric  models for univariate functional outcomes \citep{dong2024modeling}, Gaussian processes for multivariate functional responses with hierarchical mean structures \citep{li2020region}, and multilevel functional principal component analysis (FPCA) for multivariate functions \citep{campos2022multilevel, zhang2023interpretable}.  Among these, \citet{dong2024modeling} and \citet{li2020region} incorporate covariate information into the fixed effects, whereas  \citet{campos2022multilevel} and \citet{zhang2023interpretable} focus on covariance decomposition without  covariate modeling.

Beyond EEG applications, a wide range of statistical methods have been proposed to extract low-dimensional and interpretable features from  multivariate or higher-order functional data. These include singular value decomposition (SVD) for multivariate functions \citep{alam2024supervised}  and  tensors with one functional mode \citep{han2024guaranteed},  CANDECOMP/PARAFAC (CP) decomposition for tensors with one functional mode \citep{sort2024latent}, FPCA for multi-way functional data \citep{allen2013multi,chen2017multi} and  multivariate functional data \citep{happ2018multivariate}, and  latent factor Gaussian processes for two-way functional data \citep{shamshoian2022bayesian}. While these methods are powerful for dimension reduction,  they do not account for study designs with multilevel data structures. 

Existing methods for multilevel functional data  are limited to univariate or multivariate functions, and are not directly applicable to two-way functional observations. These include multilevel FPCA approaches \citep{di2009multilevel,greven2011longitudinal,zipunnikov2014longitudinal, cui2023fast}, (generalized) linear mixed-effects models \citep{guo2002functional,morris2003wavelet,morris2006wavelet,cederbaum2016functional, cui2022fast, sun2024ultra}, and (generalized) additive mixed-effects models  \citep{scheipl2015functional,Scheipl2016,volkmann2023multivariate}. Other related methods include linear mixed-effects models for matrix-valued data \citep{huang2019multilevel} and CP decomposition for multilevel tensor data  \citep{tang2020individualized}. While these methods can model smooth variation over the functional domains when applied to two-way functional data,  they  become computationally expensive for dense functional observations.

To address the limitations of existing approaches, we propose a novel Bayesian mixed-effects model for two-way functional data, motivated by time-frequency analysis in multi-condition EEG experiments. To our knowledge, this is the first approach to jointly address three key challenges: (i) the multilevel data structure from multi-subject, multi-condition designs, (ii) the two-way functional nature of  time-frequency features, and (iii) the incorporation of subject-level covariates.  The model builds on a covariate-assisted CANDECOMP/PARAFAC (CP) decomposition of the fixed effects, with a sparsity-inducing prior for automatic rank selection. Random effects are specified at two levels to capture individual variability shared across conditions and specific to each condition.  For scalable posterior inference, we generalized the efficient posterior sampling algorithm of \citet{sun2024ultra} from univariate to two-way functional mixed-effects models. 

A key advantage of our approach is its interpretability. The proposed approach identifies two-way functional patterns in the fixed effects shared across subjects and conditions, which can be decomposed into marginal functional components along the time and frequency dimensions. In the EEG context, these marginals  often carry meaningful neurophysiological interpretations, such as temporal dynamics and frequency-specific activation of neural responses.  Covariates influence how these patterns are modulated across conditions, allowing us to evaluate the impact of subject-level covariates on neural responses and to make predictions under covariate shifts.  This covariate-assisted formulation of the fixed effects can be viewed as an extension of the supervised CP decomposition \citep{lock2018supervised} to functional data. 
In addition, our model supports inference on random effects that capture both shared and condition-specific variation at the subject level.  This provides  a more complete understanding of individual differences in neural responses,  beyond those that can be explained by the observed covariates.

The remainder of the paper is organized as follows. \Cref{sec:method} introduces the proposed mixed-effects model and describes an efficient Markov chain Monte Carlo (MCMC) algorithm for posterior inference with rank selection. \Cref{sec:simulation} evaluates the proposed method through simulation studies, comparing its performance to existing alternatives. \Cref{sec:application} demonstrates the approach using EEG data from an alcoholism study, and \Cref{sec:discussion} discusses the key findings and contributions. 
 
\section{Methodology} \label{sec:method}
We consider data collected from $n$ subjects across $J$ experimental conditions. For each subject with subject-level covariates $\mathbf{x}_i \in \mathbb{R}^p$ ($i = 1, \dots, n$),  we observe two-way functions $\mathcal{Y}_{i,j}(\mathbf{x}_i): \mathcal{I}_1 \times \mathcal{I}_2 \rightarrow  \mathbb{R}$, under conditions  $j \in \mathcal{J}_i \subseteq \{1, \dots J \}$, where $\bigcup_{i=1}^n \mathcal{J}_i = \{1, \dots J\}$.  Let us denote noisy realizations of $\mathcal{Y}_{i,j}$  by  $Y_{i,j}$. We model $Y_{i,j}(t,f)$ in a  mixed-effects framework:
\begin{equation} \label{eq:2d:func:model}
Y_{i,j}(t,f) = \mathcal{A}_j(\mb{x}_i)(t,f) +  \mathcal{B}_i(t,f) + \mathcal{C}_{i,j}(t,f) + \mathcal{E}_{i,j}(t,f),   \ t \in \mathcal{I}_1, \  f \in \mathcal{I}_2,
\end{equation}
where $\mathcal{A}_j(\mb{x}_i)(t,f)$ denotes the covariate-dependent fixed effect for condition $j$,  $\mathcal{B}_i(t,f)$ captures the random effect shared across conditions for subject $i$,  $\mathcal{C}_{i,j}(t,f)$  represents the condition-specific random effect for subject $i$, and $\mathcal{E}_{i,j}(t,f)$ accounts for the random noise. In the absence of covariates, model \eqref{eq:2d:func:model} reduces to a  functional analysis of variance (FANOVA) \citep{citeulike:11611491} model for two-way functional data, where the fixed-effect $\mathcal{A}_j(t,f)$ represents the  population mean surface corresponding to experimental condition $j$.

In model \eqref{eq:2d:func:model}, we assume that the fixed and random effects ($\mathcal{A}_j(\mb{x}_i)$,  $\mathcal{B}_i$, and $\mathcal{C}_{i,j}$) are smooth functions defined over the two-way domain $\mathcal{I}_1 \times \mathcal{I}_2$.  These functions are estimated from discrete observations  of $Y_{i,j}(t, f)$  on a  grid $\mathbf{t} \times \mathbf{f}$, where $\mb{t} = (t_1, \dots, t_T)$ and  $\mb{f} = (f_1, \dots, f_F)$.  In the EEG setting, these observations correspond to measurements of neural oscillations (e.g., oscillatory power) evaluated at $T$ time points and $F$ frequency points. The noise term $\mathcal{E}_{i,j}(t,f)$ is assumed to have zero mean and is independent across $(t,f)$ and from random effects ($\mathcal{B}_i$ and   $\mathcal{C}_{i,j}$). We allow for subject-level missing conditions-- that is, not all conditions need to be observed for every subject, as is often the case in practice.

\subsection{Tensor product basis representation}
We represent the two-way functions $\mathcal{A}_j(\mb{x})$, $\mathcal{B}_i$ and $\mathcal{C}_{i,j}$ in  \eqref{eq:2d:func:model} using a tensor product basis $\Omega(t,f) = \bs{\psi}(f)  \otimes \bs{\phi}(t)  \in \mathbb{R}^K$,  constructed as the Kronecker product of marginal basis functions  $\bs{\phi}(t) = (\phi_1(t), \dots, \phi_{K_T}(t))^T \in \mathbb{R}^{K_T}$ (for the time domain)  and $\bs{\psi}(f) = (\psi_1(f), \dots, \psi_{K_F}(f))^T \in \mathbb{R}^{K_F}$ (for the frequency domain), where $K = K_TK_F$. Specifically, the fixed and random effects are expressed as:
\begin{equation} \label{eq:basis:approx}
    \mathcal{A}_j(\mb{x})(t,f)  = \mb{a}_j(\mb{x})^T\Omega(t,f), \quad \mathcal{B}_i(t,f)  = \mb{b}_i^T\Omega(t,f), \quad \mathcal{C}_{i,j}(t,f)  = \mb{c}_{i,j}^T\Omega(t,f), 
\end{equation}
where $\mb{a}_j(\mb{x}) = (a_{j,1}(\mb{x}), \dots, a_{j,K}(\mb{x}))^T$, $\mb{b}_i = (b_{i,1}, \dots, b_{i,K})^T$, and  $\mb{c}_{i,j} = (c_{i,j,1}, \dots, c_{i,j,K})^T$  are the corresponding basis coefficient vectors to be estimated from the data.   

Let $Y_{i,j}(\mb{t},\mb{f}) \in \mathbb{R}^{T \times F}$ and $\mathcal{E}_{i,j}(\mb{t},\mb{f}) \in \mathbb{R}^{T \times F}$ denote the evaluations of $Y_{i,j}(t,f)$ and $\mathcal{E}_{i,j}(t,f)$ over the grid $\mb{t} \times \mb{f}$. We define the  basis evaluation matrix as  $\mathcal{O}  = \bs{\psi}(\mb{f}) \otimes \bs{\phi}(\mb{t}) \in \mathbb{R}^{ (TF) \times  K}$, where $\bs{\phi}(\mb{t}) \in \mathbb{R}^{T \times K_T}$ and  $\bs{\psi}(\mb{f}) \in \mathbb{R}^{F \times K_F}$ correspond to the evaluations of  $\bs{\phi}(t)$ and $\bs{\psi}(f)$ over $\mb{t}$ and $\mb{f}$, respectively. 
With this representation, the discretized version of model \eqref{eq:2d:func:model} can be expressed in vectorized form as 
 \begin{equation}  \label{eq:basis:model:1}
    \mb{y}_{i,j}=   \mathcal{O} (\mb{a}_j(\mb{x}_i) + \mb{b}_i + \mb{c}_{i,j})  + \bs{\epsilon}_{i,j}, \ i = 1, \dots, n, \ j \in \mathcal{J}_i,
 \end{equation}
where $\mb{y}_{i,j} = \text{Vec}(Y_{i,j}(\mb{t},\mb{f}))$ and $\bs{\epsilon}_{i,j} = \text{Vec}(\mathcal{E}_{i,j}(\mb{t},\mb{f}))$, with $\text{Vec}(\cdot)$ denoting column-wise vectorization.

 \subsection{Model reparametrization}  
For efficient posterior inference (see \Cref{subsec:efficient}), we reparameterize \eqref{eq:basis:model:1} using  an orthogonalized basis matrix $\tilde{\mathcal{O}} =\tilde{\bs{\psi}}(\mb{f}) \otimes \tilde{\bs{\phi}}(\mb{t})$, where $\tilde{\bs{\psi}}(\mb{f}) = \bs{\psi}(\mb{f}) \mb{V}_{\psi} $ and  $\tilde{\bs{\phi}}(\mb{t}) = \bs{\phi}(\mb{t}) \mb{V}_{\phi}$ are rotated versions of the original marginal basis evaluations $\bs{\psi}(\mb{f})$ and $\bs{\phi}(\mb{t})$.  The rotated basis functions are given by $\tilde{\bs{\psi}}(f) = \mb{V}_{\psi}^T \bs{\psi}(f)$ and  $\tilde{\bs{\phi}}(t) = \mb{V}_{\phi}^T \bs{\phi}(t)$, in which the rotation matrices $\mb{V}_{\phi}$ and $\mb{V}_{\psi}$  are obtained via the SVD of the marginal basis evaluations: $\bs{\psi}(\mb{f}) = \mb{U}_{\psi} \bs{\Lambda}_{\psi}\mb{V}_{\psi}^T$ and $\bs{\phi}(\mb{t}) = \mb{U}_{\phi} \bs{\Lambda}_{\phi} \mb{V}_{\phi}^T$. We assume that $\bs{\psi}(\mb{f})$ and $\bs{\phi}(\mb{t})$ have full column rank. 
Under this assumption, $\tilde{\mathcal{O}}^T\tilde{\mathcal{O}} =  \operatorname{Diag}\left(d_{(1)}, \dots, d_{(K)}\right)$, where $d_{(k)} > 0$ for $k = 1, \dots, K$. 
We then reparameterize \eqref{eq:basis:model:1} as  
\begin{equation}  \label{eq:basis:model:2}
    \mb{y}_{i,j}=   \tilde{\mathcal{O}} (\bs{\alpha}_{j}(\mb{x}_i)  + \bs{\gamma}_{i}+  \bs{\omega}_{i,j})  + \bs{\epsilon}_{i,j}, \ i = 1, \dots, n, \ j \in \mathcal{J}_i,
 \end{equation} 
where the reparametrized coefficients 
 $\bs{\alpha}_{j}(\mb{x}_i) = (\mb{V}_{\psi} \otimes \mb{V}_{\phi})^{T} \mb{a}_{j}(\mb{x}_i)$, $\bs{\gamma}_{i} = (\mb{V}_{\psi} \otimes \mb{V}_{\phi})^{T} \mb{b}_{i}$, and  $\bs{\omega}_{i,j} = (\mb{V}_{\psi} \otimes \mb{V}_{\phi})^{T} \mb{c}_{i,j}$ are  rotated versions of $\mb{a}_{j}(\mb{x}_i)$, $\mb{b}_{i}$, and $\mb{c}_{i}$ in   \eqref{eq:basis:model:1}. This reparameterization, following \cite{sun2024ultra}, allows the random effect functions $\mathcal{B}_i(t,f)$ and $\mathcal{C}_{i,j}(t,f)$ in \eqref{eq:basis:approx} to be flexibly modeled. A posterior sampling framework is introduced  in \Cref{subsec:efficient} for $\bs{\alpha}_j(\mb{x})$, $\bs{\gamma}_i$ and $\bs{\omega}_{i,j}$ in  \eqref{eq:basis:model:2}, which can be transformed back to the original coefficients $\mb{a}_j(\mb{x})$,  $\mb{b}_i$, and $\mb{c}_{i,j}$ in \eqref{eq:basis:approx}.

\subsection{Covariate-dependent decomposition of fixed effects}
To yield a parsimonious and interpretable representation of $\mathcal{A}_j(\mb{x})$ in model \eqref{eq:2d:func:model},
we represent the fixed effects coefficients $\bs{\alpha}_j(\mb{x})$ in \eqref{eq:basis:model:2} using a novel covariate-dependent variant of the CANDECOMP/PARAFAC (CP) decomposition \citep{kolda2009tensor}.   Specifically, we define $\bs{\alpha}_j(\mb{x}) = \text{Vec} \left(\tilde{\mb{A}}_j(\mb{x})\right) \in \mathbb{R}^K$, where
 $\tilde{\mb{A}}_j(\mb{x}) \in \mathbb{R}^{K_T \times K_F}$ admits a rank-R CP decomposition:  \begin{equation} \label{eq:cp:decomposition}
 \tilde{\mb{A}}_j(\mb{x}) = \sum_{r = 1}^R \lambda_{j, r} (\mb{x})\mb{u}_{r} \mb{v}_{r}^T. 
 \end{equation}
This decomposition expresses $\tilde{\mb{A}}_j(\mb{x})$ as a covariate-weighted sum of  rank-one matrix components $\mb{u}_r \mb{v}_r^T, r=1,\ldots, R$, with $R \leq \min(K_T, K_F)$. The factor loading vectors 
 $\mb{u}_r \in \mathbb{R}^{K_T}$ and $\mb{v}_r \in \mathbb{R}^{K_F}$  represent marginal basis coefficients for the time and frequency dimensions, respectively. The  covariate-dependent and component-specific weights $\lambda_{j, r} (\mb{x})$ are modeled as linear functions of $\mb{x} \in \mathbb{R}^p$: $\lambda_{j, r} (\mb{x})  =\bs{\delta}_{j,r}^T\mb{x}$, where  $\bs{\delta}_{j,r}\in \mathbb{R}^p$. To ensure identifiability of \eqref{eq:cp:decomposition}, we assume that the factor matrices  $\mb{U} = (\mb{u}_1, \dots \mb{u}_R) \in \mathcal{V}_{R}\left(\mathbb{R}^{K_T}\right)$ and  $\mb{V} = (\mb{v}_1, \dots \mb{v}_R) \in \mathcal{V}_{R}\left(\mathbb{R}^{K_F}\right)$ lie on Stiefel manifolds, satisfying  $\mb{U}^T\mb{U}  =  \mathbf{I}_{R}$ and $\mb{V}^T\mb{V}  =  \mathbf{I}_{R}$, where $\mb{I}_R$ denotes the $R \times R$ identity matrix. When the rank $R$ is small, \eqref{eq:cp:decomposition} provides a low-rank representation of $\tilde{\mb{A}}_j(\mb{x})$, effectively regularizing the fixed-effect basis coefficients  to prevent overfitting (see Section~\ref{subsubsec:covariate} for an automatic rank selection as part of posterior inference).

Building on \eqref{eq:cp:decomposition}, $\mathcal{A}_j(\mb{x})$ in the two-way functional domain
(\ref{eq:basis:approx}) can be expressed as
\begin{align} \label{eq:2way:decomp}
\mathcal{A}_j(\mb{x})(t,f)  &=
\sum_{r = 1}^R \lambda_{j,r}(\mb{x}) \phi_r^{\ast}(t) \psi_r^{\ast}(f), 
\end{align}
where 
$\phi_r^{\ast}(t) =  \mb{u}_{r}^T  \tilde{\bs{\phi}}(t)$ and $\psi_r^{\ast}(f) =   \mb{v}_{r}^T \tilde{\bs{\psi}}(f)$.
$\phi_r^{\ast}(t)$ and $\psi_r^{\ast}(f)$ can be interpreted as marginal ``principal'' functions along the time and frequency domains, respectively, which often carry neurologically meaningful interpretations in EEG studies (see \Cref{sec:application} for illustration).  The products $\phi_r^{\ast}(t)\psi_r^{\ast}(f)$ define two-way ``base'' patterns that are shared across conditions and act as building blocks for
 $\mathcal{A}_j(\mb{x})(t,f)$. Each ``base'' pattern is scaled by a condition- and covariate-dependent weight  $\lambda_{j,r}(\mb{x})$,  allowing subject characteristics to modulate the expression of this ``base'' pattern. When  $\mb{x}_i = 1$ for all  $i$, the fixed effect \eqref{eq:2way:decomp} reduces to the condition-specific population mean surface.

 \subsection{Efficient posterior inference}  \label{subsec:efficient}
We develop a fully Bayesian inference procedure for model \eqref{eq:basis:model:2}, enabling estimation of the fixed effect $\mathcal{A}_j(\mb{x})$ and random effects $\mathcal{B}_i$ and $\mathcal{C}_{i,j}$.  Leveraging the CP decomposition in  \eqref{eq:2way:decomp}, our approach  allows inference of  $\phi_r^{\ast}(t)$ and $\psi_r^{\ast}(f)$, as well as the weights $\lambda_{j,r}(\mb{x})$.  For computational efficiency, we build on the inference approach proposed by \citet{sun2024ultra}, adapting their strategy for sampling the random effects while introducing a new sampling procedure tailored to the structured fixed effects.

Throughout this paper, the subscript $(l)$ denotes quantities (scalars, vectors, or matrices) associated with the $l$-th tensor product basis function. Accordingly,  $\gamma_{i,(l)}$,  $\omega_{i,j,(l)}$, and $\epsilon_{i,j,(l)}$ represent the $l$-th elements of $\bs{\gamma}_i$,  $\bs{\omega}_{i,j}$, and  $\bs{\epsilon}_{i,j}$, respectively. We assign the following priors to the random components:
\begin{equation} \label{eq:prior:effects}
    \gamma_{i,(l)}|\sigma^2_{\gamma} \sim N(0, \sigma^2_{\gamma}), \   \  \omega_{i,j,(l)}|\sigma^2_{\omega_i} \sim N(0, \sigma^2_{\omega_i}), \ \ \epsilon_{i,j,(l)}|\sigma^2_{\epsilon} \sim N(0, \sigma^2_{\epsilon}), 
\end{equation}
with the following hyperpriors on the variance parameters:
\begin{equation}\label{eq:prior:var}
   \sigma^{2}_{\gamma} \sim  \text{IG}(a_\gamma,b_\gamma), \   \   \sigma^{2}_{\omega_i} \sim  \text{IG}(a_{\omega},b_{\omega}), \ \ p(\sigma_{\epsilon}^2) \propto 1/\sigma_{\epsilon}^2, 
\end{equation} 
with $\text{IG}(a,b)$  denoting an inverse Gamma distribution with shape parameter $a$ and scale parameter $b$. For $\sigma^2_{\omega}$, we also consider a simplified setting with homogeneous variances across subjects,  i.e., $\sigma^2_{\omega_i} = \sigma^2_{\omega}$ for all $i$.  The corresponding changes to the posterior sampling procedure  are supported in our implementation and detailed in Supplementary Material A. Priors for the fixed effects are introduced in Sections \ref{subsubsec:factor} and \ref{subsubsec:covariate}.
Identifiability of model \eqref{eq:basis:model:2}, under the CP representation in \eqref{eq:cp:decomposition} and the priors in \eqref{eq:prior:effects}, is established in Supplementary Material B.

\textit{Notation.} We denote the observed data as $\mathcal{D} =  \{\mb{X}_{1:n}, \mb{Y}_{1:n}\}$,  where $\mb{Y}_{1:n}= \{\{\mb{y}_{i,j}\}_{j \in \mathcal{J}_i}\}_{i=1}^n$ and $\mb{X}_{1:n}= \{\mb{x}_1, \dots, \mb{x}_n\}$.  The variance components are collectively denoted by $\mathcal{S}  = \{\sigma_{\gamma}^2,  \sigma_{\epsilon}^2, \bs{\sigma}_{\omega}^2\}$, where $\bs{\sigma}_{\omega}^2 = \{\sigma_{\omega_1}^2, \dots, \sigma_{\omega_n}^2\}$.  We denote the sets of fixed and random effect coefficients as follows:
$\bs{\alpha}_{1:J} = \{\bs{\alpha}_1, \dots, \bs{\alpha}_J \}$, $\bs{\gamma}_{1:n} = \{\bs{\gamma}_1, \dots, \bs{\gamma}_n \}$, and $\bs{\omega}_{1:n} = \{\bs{\omega}_1, \dots, \bs{\omega}_n \}$, where $\bs{\omega}_i = \{\bs{\omega}_{i,j}, \  j \in \mathcal{J}_i \}$. Similarly, we denote the sets of random noise as $\bs{\epsilon}_{1:n} = \{\bs{\epsilon}_1, \dots, \bs{\epsilon}_n \}$, where   $\bs{\epsilon}_i = \{\bs{\epsilon}_{i,j}, \  j \in \mathcal{J}_i \}$. We also introduce notation for the random coefficients associated with the $l$-th tensor product basis function:  $\bs{\gamma}_{(l)} = (\gamma_{1,(l)}, \dots, \gamma_{n,(l)})$ and $\bs{\omega}_{(l)} = (\bs{\omega}_{1,(l)}^T, \dots, \bs{\omega}_{n,(l)}^T)^T$, where $\bs{\omega}_{i,(l)} = (\omega_{i,j, (l)})_{j \in \mathcal{J}_i}$. The set of covariate coefficients is denoted by   
$\bs{\delta}_{1:J, 1:R} = \{\{\bs{\delta}_{j,r}\}_{j = 1}^J\}_{r=1}^R$. We use $\mb{0}_m$ and $\mb{1}_m$ to denote the $m$-dimensional column vectors of zeros and ones, respectively, and $\mb{I}_m$ to denote the $m \times m$ identity matrix.

The posterior sampling procedure is summarized in \Cref{alg:1}. Inference is conducted via a two-block Gibbs sampler (in which the subscript $t$ denotes the sample at the $t$-th iteration): \textbf{Block 1}: sample the fixed and random effect coefficients    $(\bs{\alpha}_{1:J}, \bs{\gamma}_{1:n}, \bs{\omega}_{1:n} \mid \mathcal{S}_t, \mathcal{D})$;  and \textbf{Block~2}:  sample the variance components $(\mathcal{S}\mid \bs{\alpha}_{1:J,t}, \bs{\gamma}_{1:n,t}, \bs{\omega}_{1:n,t}, \mathcal{D})$. 

 \subsubsection{Posterior sampling of the variance components}  
 
We begin by describing \textbf{Block~2}, which involves standard conjugate updates. Sampling steps for \textbf{Block~1} are detailed  in Sections \ref{subset:random}, \ref{subsubsec:factor} and \ref{subsubsec:covariate}. 

Let $J_i = |\mathcal{J}_i|$ denote the number of observed conditions for subject $i$, where $|\cdot|$ denotes set cardinality, and  define $J' = \sum_{i=1}^n J_i$. In \textbf{Block 2}, we sample the variance components $\mathcal{S}$ from their conditional posterior:
\begin{align*}
p(\mathcal{S}|\bs{\alpha}_{1:J}, \bs{\gamma}_{1:n}, \bs{\omega}_{1:n}, \mathcal{D}) 
 &\propto  p(\mathcal{S})p(\bs{\alpha}_{1:J}, \bs{\gamma}_{1:n}, \bs{\omega}_{1:n}|\mathcal{S})p(\mb{Y}_{1:n}| \bs{\alpha}_{1:J}, \bs{\gamma}_{1:n}, \bs{\omega}_{1:n}, \mathcal{S}, \mb{X}_{1:n}).
\end{align*}
This yields the conjugate updates below, with derivations provided in Supplementary Material C:
\begin{align}
   & \sigma_{\gamma}^{2}|\bs{\alpha}_{1:J}, \bs{\gamma}_{1:n},  \bs{\omega}_{1:n},   \mathcal{D}  \sim \text{IG} \left(a_{\gamma} + \frac{1}{2} nK, b_{\gamma}+ \frac{1}{2}  \sum_{i=1}^n \sum_{l=1}^K  \gamma_{i,(l)}^2 \right), \label{eq:block2:line2}  \\
   & \sigma_{\omega_i}^{2}|\bs{\alpha}_{1:J}, \bs{\gamma}_{1:n},  \bs{\omega}_{1:n},   \mathcal{D}  \sim \text{IG} \left(a_{\omega} + \frac{1}{2} J_i K, b_{\omega}+ \frac{1}{2} \sum_{j \in \mathcal{J}_i} \sum_{l=1}^K  \omega_{i,j, (l)}^2 \right), \label{eq:block2:line3}\\
       &\sigma_{\epsilon}^{2}|\bs{\alpha}_{1:J}, \bs{\gamma}_{1:n},  \bs{\omega}_{1:n},   \mathcal{D} \sim \text{IG} \left(\frac{1}{2} TF J', \frac{1}{2} \sum_{i=1}^n \sum_{j \in \mathcal{J}_i} ||\mb{y}_{i,j} - \tilde{\mathcal{O}} \bs{\beta}_{i,j}(\mb{x}_i)||_2^2 \right),   \label{eq:block2:line1}
\end{align}
where $ \bs{\beta}_{i,j}(\mb{x}_i) =  \bs{\alpha}_{j}(\mb{x}_i)  + \bs{\gamma}_{i}+  \bs{\omega}_{i,j} $.  

\SetKwFor{If}{If}{}{}
\SetKwFor{Else}{Else}{}{EndIf}
\SetKwProg{Input}{Input:}{}{}
\SetKwProg{Return}{Return:}{}{}
\SetKwProg{Initialization}{Initialize:}{}{}
\SetKwProg{Parameters}{Parameters:}{}{}
\SetKwFor{For}{For}{do}{EndFor}
\SetKwProg{DefData}{Data:}{}{}

\begin{algorithm} \caption{Efficient posterior sampling algorithm} \label{alg:1}
\DefData{$\mathcal{D} =  \{\mb{X}_{1:n}, \mb{Y}_{1:n}\}$}{}
\textbf{For} {$t = 1, \dots, T$} \textbf{do}

\hspace{1em}
\textbf{Block 1}: Sample the fixed and random effects coefficients $(\bs{\alpha}_{1:J}, \bs{\gamma}_{1:n},  \bs{\omega}_{1:n}| \mathcal{S}_t , \mathcal{D})$

\vspace{0.3em}

\hspace{1em}
 (a) Sample $(\bs{\alpha}_{1:J} \big|  \mathcal{S}_{t-1}, \mathcal{D})$  \algorithmiccomment{See  Sections \ref{subsubsec:factor} and \ref{subsubsec:covariate}} \\

\hspace{1em}
(b) \textbf{For} {$l = 1, \dots, K$} \textbf{do} \\
\hspace{3em} Sample $(\bs{\gamma}_{(l)} \big|  \bs{\alpha}_{1:J, t},   \mathcal{S}_{t-1}, \mathcal{D})$  \algorithmiccomment{See  \Cref{subset:random}}   \\ 
\hspace{2em} \textbf{ EndFor}

\hspace{1em}
(c) \textbf{For} {$l = 1, \dots, K$} \textbf{do} \\
\hspace{3em} Sample $ \big(\bs{\omega}_{(l)} \big| \bs{\alpha}_{1:J, t},  \bs{\gamma}_{1:n, t},   \mathcal{S}_{t-1}, \mathcal{D} \big)$  \algorithmiccomment{See  \Cref{subset:random}} \\
\hspace{2em} \textbf{EndFor}

\hspace{1em}
\textbf{Block 2}: Sample the variance components $(\mathcal{S}|\bs{\alpha}_{1:J,t}, \bs{\gamma}_{1:n,t},  \bs{\omega}_{1:n,t},   \mathcal{D})$

\hspace{1em}
(a) Sample  $(\sigma_{\gamma}^2| \bs{\alpha}_{1:J,t}, \bs{\gamma}_{1:n,t},  \bs{\omega}_{1:n,t},   \mathcal{D})$  from \eqref{eq:block2:line2} \\

\hspace{1em}
(b) Sample  $(\sigma_{\omega_i}^2| \bs{\alpha}_{1:J,t}, \bs{\gamma}_{1:n,t},  \bs{\omega}_{1:n,t},   \mathcal{D})$ from \eqref{eq:block2:line3}, for $i = 1, \dots, n$    \\

\hspace{1em}
(c) Sample $(\sigma_{\epsilon}^2| \bs{\alpha}_{1:J,t}, \bs{\gamma}_{1:n,t},  \bs{\omega}_{1:n,t},   \mathcal{D})$ from \eqref{eq:block2:line1} \\

\textbf{EndFor}

\Return{$\{\bs{\alpha}_{1:J,t}\}_{t=1}^T$,  $\{\bs{\gamma}_{1:n,t}\}_{t=1}^T$,  $\{\bs{\omega}_{1:n,t}\}_{t=1}^T$, $\{\mathcal{S}_t\}_{t=1}^T$}{}
\end{algorithm}

\subsubsection{Posterior sampling of random effects}  \label{subset:random}
We adopt the projection approach of \citet{sun2024ultra} to simplify \eqref{eq:basis:model:2} for efficient posterior sampling.  For each $\mb{y}_{i,j} \in \mathbb{R}^{TF}$, we define the  projection 
$\tilde{\mb{y}}_{i,j} =  (\tilde{\mathcal{O}}^T\tilde{\mathcal{O}})^{-1} \tilde{\mathcal{O}}^T\mb{y}_{i,j} \in \mathbb{R}^{K}$, which yields the reduced model:
\begin{align}  \label{eq:basis:model:3}
    \tilde{\mb{y}}_{i,j} &=  \bs{\alpha}_{j}(\mb{x}_i)  +  \bs{\gamma}_{i}  + \bs{\omega}_{i,j}  + \tilde{\bs{\epsilon}}_{i, j}, \bigskip   \  
\end{align}
where $\tilde{\bs{\epsilon}}_{i, j} \sim N\left(\mb{0}_K,   \ \sigma_{\epsilon}^2 \mb{D}^{-1} \right)$, with $\mb{D} =\tilde{\mathcal{O}}^T\tilde{\mathcal{O}}$. 
The likelihood of \eqref{eq:basis:model:3} is proportional to that of  \eqref{eq:basis:model:2}, since 
 $\exp \left( - || \mb{y}_{i,j}  -\tilde{\mathcal{O}} \bs{\beta}_{i,j}(\mb{x}_i) ||_2^2/ (2\sigma_{\epsilon}^2) \right) \propto \exp \left( - || \sqrt{\mb{D}} \left(\tilde{\mb{y}}_{i,j} - \bs{\beta}_{i,j}(\mb{x}_i) \right) ||_2^2/(2\sigma_{\epsilon}^2) \right)$ (a complete derivation is provided in Supplementary Material C). 

 We reformulate \eqref{eq:basis:model:3} by stacking parameters (across conditions and then subjects) associated with each ($l = 1, \dots K$) tensor product basis function,  yielding the following model:
\begin{align}  \label{eq:basis:model:4}
    \tilde{\mb{y}}_{(l)} &=   \bs{\alpha}_{(l)} + \mb{Z}{\bs{\gamma}_{(l)}} +  \bs{\omega}_{(l)} + \tilde{\bs{\epsilon}}_{(l)},  \ \bs{\epsilon}_{(l)} \sim N\left(\mb{0}_{J'},  \mb{\Sigma}_{\tilde{\epsilon}_{(l)}}\right),
\end{align}
where  $\tilde{\mb{y}}_{(l)} = (\tilde{\mb{y}}_{1,(l)}^T , \dots, \tilde{\mb{y}}_{n,(l)}^T)^T$, $\tilde{\mb{y}}_{i,(l)} = (\tilde{y}_{i,j, (l)})_{j \in \mathcal{J}_i}$ (which represents vertical stacking across conditions $j \in \mathcal{J}_i$); $\bs{\alpha}_{(l)} =  (\bs{\alpha}_{1,(l)}^T , \dots, \bs{\alpha}_{n,(l)}^T)^T$, $\bs{\alpha}_{i,(l)} = (\alpha_{j,(l)}(\mb{x}_i))_{j \in \mathcal{J}_i}$;  $\mb{Z} = \text{BlockDiag}(\mb{1}_{J_i},  \dots, \mb{1}_{J_n}) \in \mathbb{R}^{J' \times n}$; $\tilde{\bs{\epsilon}}_{(l)} = (\tilde{\bs{\epsilon}}_{1,(l)}^T, \dots, \tilde{\bs{\epsilon}}_{n,(l)}^T)^T$,  $\tilde{\bs{\epsilon}}_{i, (l)} = (\tilde{\epsilon}_{i,j, (l)})_{j \in \mathcal{J}_i}$; $\mb{\Sigma}_{\tilde{\epsilon}_{(l)}} =\sigma_{\tilde{\epsilon}_{(l)}}^2 \mb{I}_{J'}$, where $\sigma_{\tilde{\epsilon}_{(l)}}^2 = \sigma_{\epsilon}^2/d_{(l)}$; and $\bs{\gamma}_{(l)}$ and $\bs{\omega}_{(l)}$ are defined in \Cref{subsec:efficient}. We assume, without loss of generality, that the indices  $j \in \mathcal{J}_i$ are sorted in ascending order.

Based on model \eqref{eq:basis:model:4} and priors specified in \eqref{eq:prior:effects}, the posterior distributions of $\bs{\omega}$ and $\bs{\gamma}$ factorize over the basis indices $l = 1, \dots K$, allowing independent sampling for each $l$:   
\begin{align}
p(\bs{\omega}| \bs{\alpha}, \bs{\gamma},  \mathcal{S}, \mathcal{D}) &= \prod_{l=1}^K  p(\bs{\omega}_{(l)}| \bs{\sigma}_{\omega}^2) p(\tilde{\mb{y}}_{(l)}|\bs{\alpha}_{(l)}, \bs{\gamma}_{(l)}, \bs{\omega}_{(l)},  \sigma_{\tilde{\epsilon}_{(l)}}^2), \label{eq:posterior:1} \\
p(\bs{\gamma}|\bs{\alpha}, \mathcal{S}, \mathcal{D})  
&= \prod_{l=1}^K  p(\bs{\gamma}_{(l)}| \sigma_{\gamma}^2) p(\tilde{\mb{y}}_{(l)}|\bs{\alpha}_{(l)}, \bs{\gamma}_{(l)},  \bs{\sigma}_{\omega}^2, \sigma_{\tilde{\epsilon}_{(l)}}^2).
\label{eq:posterior:2}
\end{align}
The likelihood terms in \eqref{eq:posterior:1} and \eqref{eq:posterior:2}  follow directly from \eqref{eq:basis:model:4} and its marginalization over $\bs{\omega}_{(l)}$. For $l = 1, \dots, K$, the resulting posteriors are
\begin{align} 
    \bs{\omega}_{(l)}| \bs{\alpha}_{(l)}, \bs{\gamma}_{(l)},   \mathcal{S}, \mathcal{D}   &  \sim N(\mb{C}_{\bs{\omega}_{(l)}} \mb{g}_{\bs{\omega}_{(l)}} , \mb{C}_{\bs{\omega}_{(l)}}), \label{eq:posterior:random1}\\  \bs{\gamma}_{(l)}| \bs{\alpha}_{(l)}, \mathcal{S}, \mathcal{D}  &\sim N( \mb{C}_{\gamma_{(l)}} \mb{g}_{\gamma_{(l)}}, \mb{C}_{\gamma_{(l)}}), \label{eq:posterior:random2}
\end{align}
where $\mb{C}_{\bs{\omega}_{(l)}}  = (\bs{\Sigma}_{\omega}^{-1} + \bs{\Sigma}_{\epsilon_{(l)}}^{-1})^{-1} \in \mathbb{R}^{J' \times J'}$,  $\mb{g}_{\bs{\omega}_{(l)}} = \bs{\Sigma}_{\epsilon_{(l)}}^{-1}
\left(\tilde{\mb{y}}_{(l)} -  \bs{\alpha}_{(l)} - \mb{Z}\gamma_{(l)}\right) \in \mathbb{R}^{J'}$,  $\mb{C}_{\gamma_{(l)}} =  (\bs{\Sigma}_{\gamma}^{-1} + \mb{Z}^T (\bs{\Sigma}_{\omega} + \bs{\Sigma}_{\epsilon_{(l)}})^{-1} \mb{Z})^{-1} \in \mathbb{R}^{n\times n}$,  $\mb{g}_{\gamma_{(l)}} = \mb{Z}^T (\bs{\Sigma}_{\omega} + \bs{\Sigma}_{\epsilon_{(l)}})^{-1} (\tilde{\mb{y}}_{(l)} - \bs{\alpha}_{(l)}) \in \mathbb{R}^n$, $\bs{\Sigma}_{\omega} = \text{BlockDiag}(\sigma_{\omega_1}^2 \mb{I}_{J_1}, \dots, \sigma_{\omega_n}^2 \mb{I}_{J_n}) \in \mathbb{R}^{J' \times J'}$, and $\bs{\Sigma}_{\gamma} = \sigma_{\gamma}^2 \mb{I}_n$ (derivations are provided in Supplementary Material D).  In \eqref{eq:posterior:random1} and \eqref{eq:posterior:random2}, the covariance matrices $\mb{C}_{\omega_{(l)}}$ and $\mb{C}_{\gamma_{(l)}}$ are diagonal, which enables fully parallel and elementwise posterior sampling of $\bs{\omega}_{(l)}$ and $\bs{\gamma}_{(l)}$.  In particular, the diagonality of $\mb{C}_{\gamma_{(l)}}$  follows from  $$\mb{Z}^T (\bs{\Sigma}_{\omega} + \bs{\Sigma}_{\epsilon_{(l)}})^{-1} \mb{Z} = \text{Diag}\left( J_1 \left(\sigma_{\omega_1}^2+ \sigma_{\epsilon}^2/d_{(l)}\right)^{-1}, \dots, J_n \left(\sigma_{\omega_n}^2+ \sigma_{\epsilon}^2/d_{(l)} \right)^{-1} \right).$$



\subsubsection{Posterior sampling of fixed effect factor matrices} \label{subsubsec:factor}
We use a Gibbs sampling procedure for posterior inference of the factor matrices $\mb{U}$ and $\mb{V}$ in the  CP representation \eqref{eq:cp:decomposition}. This  procedure effectively addresses two key challenges:  enforcing the orthonormality constraints on $\mb{U}$ and $\mb{V}$, and selecting an appropriate rank $R$ as described in \Cref{subsubsec:covariate}.  

Let $\mb{u}_{-r} = \{\mb{u}_{r'}\}_{r' \neq r}$ and $\mb{v}_{-r} = \{\mb{v}_{r'}\}_{r' \neq r}$. We introduce the posterior sampling of $(\mb{u}_r|\mb{u}_{-r}, \mb{V}, \bs{\delta}_{1:J, 1:R} , \mathcal{S}, \mathcal{D})$ below, with an analogous derivation for  $(\mb{v}_r|\mb{v}_{-r}, \mb{U}, \bs{\delta}_{1:J, 1:R}  \mathcal{S}, \mathcal{D})$ provided in Supplementary Material D.

We consider a Gibbs sampler that updates the columns of $\mb{U}$ and $\mb{V}$ in the order $\mb{u}_{1} \rightarrow  \mb{v}_{1}  \rightarrow \mb{u}_{2} \rightarrow  \mb{v}_{2} \cdots   \rightarrow  \mb{u}_{R} \rightarrow  \mb{v}_{R}$.  Due to the orthonormality constraints, each  $\mb{u}_r$ must lie in the orthogonal complement of the space spanned by $\mb{u}_{-r}$, denoted by $\text{Span}(\mb{u}_{-r})^{\perp}$.  We construct an orthonormal basis for this orthogonal complement  space, denoted by $\mb{B}_{r} \in \mathbb{R}^{K_T \times \tilde{K}_T}$, where $\tilde{K}_T = K_T - R + 1$ represents the basis dimension excluding the $R - 1$ directions spanned by $\mb{u}_{-r}$, and  represent  $\mb{u}_{r} = \mb{B}_{r}\bs{\theta}_{r},$ where $\bs{\theta}_{r} \in \mathbb{S}^{\tilde{K}_T-1}$ lies on the unit sphere $\mathbb{S}^{\tilde{K}_T-1}  = \{ \bs{\theta} \in \mathbb{R}^{\tilde{K}_T}, ||\bs{\theta}||_2 = 1 \}$, ensuring $||\mb{u}_r||_2 = 1$ under the Euclidean norm $||\cdot||_2$.  The orthonormal basis $\mb{B}_{r}$ can be constructed via the QR decomposition of  $\mb{u}_{-r}$, through the Gram-Schmidt process \citep{stewart1973introduction}.  Posterior sampling is conducted on $\bs{\theta}_{r}$. 

We define the ``partial residuals'' for subject $i$ under condition $j$ as  
$\mb{q}_{-r, i, j} = \tilde{\mb{y}}_{i,j}- \sum_{r'\neq r}  ( \bs{\delta}_{j,r'}^T\mb{x}_i) (\mb{v}_{r'} \otimes \mb{u}_{r'}) \in \mathbb{R}^{K}.$ Stacking these vectors across $J_i$ conditions, we obtain  
$\mb{q}_{-r, i} =(\mb{q}_{-r, i, j}^T )_{j \in J_i} \in \mathbb{R}^{J_iK}$. We also  define   $\tilde{\mb{v}}_{r, i} = \tilde{\mb{\lambda}}_r(\mb{x}_i) \otimes \mb{v}_{r} \in \mathbb{R}^{J_iK_F}$, where $\tilde{\mb{\lambda}}_r(\mb{x}_i) = (\bs{\delta}_{j,r}^T\mb{x}_i)_{j \in \mathcal{J}_i}$ denotes the weights for subject $i$.  The stacked random components, including random effects and  noise are given by $\mb{q}_{-r, i}  - \tilde{\mb{v}}_{r, i} \otimes (\mb{B}_{ r}\bs{\theta}_{r})$,   with associated covariance 
\begin{equation} \label{eq:cov}
  \bs{\Sigma}_i =   \sigma_{\gamma}^2 \mb{1}_{J_iK}\mb{1}_{J_iK}^T +  \sigma_{\omega_i}^2 \mb{I}_{J_iK} + \sigma_{\epsilon}^2  \mb{I}_{J_i} \otimes  \mb{D}^{-1} \in \mathbb{R}^{(J_iK) \times (J_iK)}.  
\end{equation}
Let $\tilde{\mb{v}}_{r, 1:n} = \{\tilde{\mb{v}}_{r, i}\}_{i=1}^n$ and $\mb{q}_{-r, 1:n} = \{\mb{q}_{-r, i}\}_{i=1}^n$. Given a prior $p(\bs{\theta}_r)$ over  $\bs{\theta}_r \in   \mathbb{S}^{\tilde{K}_T-1}$, the posterior distribution of $\bs{\theta}_r$  is given by
\begin{align} 
 p \left( \bs{\theta}_{r}| \mb{v}_{r}, \tilde{\mb{v}}_{r, 1:n}, \mb{q}_{-r, 1:n},  \mathcal{S} \right)  
&\propto p(\bs{\theta}_r)\text{exp} \left(-\frac{1}{2}  \left(\sum_{i=1}^n \left(\mb{q}_{-r, i} -  \tilde{\mb{v}}_{r, i} \otimes (\mb{B}_{ r}\bs{\theta}_{r}) \right)^T   \bs{\Sigma}^{-1}_i \left(\mb{q}_{-r, i} - \tilde{\mb{v}}_{r, i} \otimes (\mb{B}_{ r}\bs{\theta}_{r}) \right) \right)\right)  \nonumber \\
&\propto  p(\bs{\theta}_r) \text{exp} \left(-\frac{1}{2}\bs{\theta}_{r}^T \left(
 \sum_{i=1}^n \mb{B}_{r}^T \mb{H}_{r, i} \mb{B}_{r} \right)\bs{\theta}_{r} +   \bs{\theta}_{r}^T \left(\mb{B}_{r}^T \sum_{i=1}^n \mb{w}_{r, i}\right)  \right), \label{eq:pos:u}
\end{align}
where $\mb{H}_{r, i}  = \sum_{l_1,l_2} \tilde{v}_{r,i, l_1} \tilde{v}_{r,i, l_2} \bs{\Sigma}_{i, [l_1,l_2]}^{-1}   \in \mathbb{R}^{K_T \times K_T}$ and  $\mb{w}_{r, i} = \sum_{l_1,l_2} \tilde{v}_{r, i, l_1}  \bs{\Sigma}_{i, [l_1,l_2]}^{-1} \mb{q}_{-r, i, [l_2]}  \in \mathbb{R}^{K_T}$, with the summations taken over $l_1, l_2 = 1, \dots, J_iK_F$.  Here, $\tilde{v}_{r,i, l}$ denotes the $l$-th element of $\tilde{\mb{v}}_{r,i}$; $\bs{\Sigma}_i^{-1}$ is treated as a block matrix with $(J_iK_F) \times (J_iK_F)$ blocks, each of size $K_T \times K_T$, with $\bs{\Sigma}_{i,[l_1,l_2]}^{-1} \in \mathbb{R}^{K_T \times K_T}$ denoting the $(l_1,l_2)$-th block; $\mb{q}_{-r, i} $ is treated as a block vector of $J_iK_F$ blocks, each of size $K_T$, with $\mb{q}_{-r, i, [l_2]} \in \mathbb{R}^{K_T}$ denoting the $l_2$-th block. The derivation of  \eqref{eq:pos:u} is provided in Supplementary Material D. 

We consider two prior specifications for $\mb{U} \in \mathcal{V}_{R}\left(\mathbb{R}^{K_T}\right)$, an uninformative (uniform) prior and a matrix von Mises–Fisher prior.  Under the uniform prior 
$p_{\mb{U}}(\mb{U}) \propto 1$, the  conditional prior on $\mb{u}_r|\mb{u}_{-r}$ is uniform in the subspace $\mb{u}_r \in \text{Span}(\mb{u}_{-r})^{\perp}$,  corresponding to  $p(\bs{\theta}_r) \propto 1$ for   $\bs{\theta}_r \in \mathbb{S}^{\tilde{K}_T-1}$ \citep{hoff2007model}.   From \eqref{eq:pos:u}, the resulting posterior is
\begin{equation} \label{eq:fisher:bin}
    \bs{\theta}_{r}| \mb{v}_{r}, \tilde{\mb{v}}_{r, 1:n}, \mb{q}_{-r, 1:n},  \mathcal{S}  \sim \text{Fisher-Bingham}(\mb{g}_{\mb{u}_{r}} , \mb{Q}_{\mb{u}_{r}}),
\end{equation}
where $\mb{g}_{\mb{u}_{r}} =  \mb{B}_{r}^T  \sum_{i=1}^n \mb{w}_{r, i}$ and $\mb{Q}_{\mb{u}_{r}}  =    \mb{B}_{r}^T (\sum_{i=1}^n  \mb{H}_{r, i})\mb{B}_{r}$. The Fisher–Bingham distribution with parameters 
$(\bs{\mu}_0, \mb{S}_0)$ has density $f_{\text{FB}}(\bs{\theta}) \propto \text{exp}(\bs{\mu}_0^T\bs{\theta} - \bs{\theta}^T\mb{S}_0\bs{\theta}/2)$ \citep{hoff2009simulation}, and samples can be efficiently drawn  using the method of \citet{kent2018new}. Alternatively,  prior information about the ``principal'' functions can be incorporated through a matrix von Mises–Fisher (MvMF) prior on $\mb{U}$ with directional preference parameter $\mb{F} \in \mathcal{V}_R(\mathbb{R}^{K_T})$ and concentration parameter $\nu >0$. The MvMF prior takes the form $f_{\text{MvMF}}(\mb{U}) \propto \text{exp}{( \nu \text{trace}( \mb{F}^T\mb{U}))} = \text{exp}\left( \nu {\sum_{r=1}^R \mb{u}_r^T\mb{f}_r}\right)$, where $\mb{f}_r$ denotes the $r$-th column of $\mb{F}$, leading to an exponential prior $p(\bs{\theta}_r) \propto  \text{exp}\left(\nu {\bs{\theta}_r^T\mb{B}_r^T\mb{f}_r}\right)$  for $\bs{\theta}_r \in \mathbb{S}^{\tilde{K}_T-1}$ 
 \citep{hoff2009simulation}. The resulting posterior distribution of $\bs{\theta}_r$ remains Fisher–Bingham as in \eqref{eq:fisher:bin}, with the same $\mb{Q}_{\mb{u}_{r}}$ and an updated  $\mb{g}_{\mb{u}_{r}} =  \mb{B}_{r}^T  \sum_{i=1}^n \mb{w}_{r, i} +  \nu \mb{B}_{r}^T \mb{f}_r$.


\subsubsection{Posterior sampling of covariate coefficients} \label{subsubsec:covariate}

For rank selection, we introduce a sparsity-inducing prior on the coefficient vectors $\bs{\delta}_{j,r}$ to shrink coefficients corresponding to redundant ranks.  Shrinkage-based rank selection has been explored using the multiplicative gamma process prior (MGPP) \citep{bhattacharya2011sparse, shamshoian2022bayesian}, which assumes decreasing importance for ranks $r = 1, \dots, R$. In contrast, our shrinkage prior imposes no ordering on rank importance, providing greater flexibility in exploring the parameter space.  This flexibility is particularly advantageous given the strong dependencies among $\mb{u}_1, \dots, \mb{u}_R$ and $\mb{v}_1, \dots, \mb{v}_R$ due to orthogonality constraints on $\mb{U}$ and $\mb{V}$, which makes additional rank-order assumptions overly restrictive and causes slow MCMC mixing.

 Motivated by the spike-and-slab LASSO (SSL) \citep{rovckova2018spike} prior originally proposed for sparse mean estimation, we specify the following hierarchical prior 
\begin{align} \label{eq:prior:delta}
    \bs{\delta}_{j,r}| \tau_r \sim  N(\mb{0}_p, \tau_r \bs{\Sigma}_{\delta_j}),
\end{align}
 where $\tau_r > 0$ controls the contribution of component $r$, and $\bs{\Sigma}_{\delta_j}$ is a user-specified  matrix encoding prior beliefs about the structure of $\bs{\delta}_{j,r}$. When $\tau_r$ takes small values,  $\bs{\delta}_{1,r}, \dots, \bs{\delta}_{J,r}$ are shrunk toward $\mb{0}_p$, suggesting negligible contribution from component $r$. To encourage low-rank structures in the CP decomposition \eqref{eq:cp:decomposition},  we impose an SSL-inspired  prior on $\tau_r$: 
\begin{align}  \label{eq:prior:ssl}
p(\tau_r|m_r) &= f_{\text{HL}, h_1}(\tau_r)^{m_r}f_{\text{HL}, h_0}(\tau_r)^{1-m_r} , \ 
p(m_r|\pi) = \text{Bernoulli}(\pi), 
\pi \sim \text{Beta}(a_{\delta}, b_{\delta}),
\end{align}
where $f_{\text{HL},h}(\tau) = \text{exp}\left(-\tau/h\right)/h$ is the density of the half-Laplace distribution with scale parameter $h$. The construction in \eqref{eq:prior:ssl} defines a mixture of two half-Laplace distributions for  $\tau_r$ with scales $h_0$ and $h_1$  ($h_0 < h_1$) and mixing proportion $\pi$.  The ``spike" component (with scale $h_0$) promotes strong shrinkage of $\tau_r$ toward zero,  pruning redundant ranks, while the ``slab" component (with scale $h_1$) permits larger values of  $\tau_r$.  The binary indicator $m_r$ determines whether rank $r$  is assigned to the      ``spike'' ($m_r = 0$) or ``slab'' ($m_r = 1$) component.  


At each MCMC iteration, after updating $\mb{U}$ and $\mb{V}$ (\Cref{subsubsec:factor}), we perform posterior sampling via Steps (1) to (4) below. The derivations are  in Supplementary Material D. For notational convenience, we define:
 $\bs{\tau}_{1:R} = \{\tau_r\}_{r =1}^{R}$, $\bs{\tau}_{-r} = \{\tau_{r'}\}_{r' \neq r}$, $\mb{m}_{1:R} = \{m_r\}_{r =1}^{R}$,  $\mb{m}_{-r} = \{m_{r'}\}_{r' \neq r}$,  $\bs{\delta}_{-(j,r)} = \{\bs{\delta}_{j', r'}\}_{(j',r') \neq (j, r)}$, and  $\bs{\Xi}_{i,r}= \mb{x}_i \left( \mb{v}_{r} \otimes \mb{u}_r \right)^T$. 

\begin{itemize}
    \item \textbf{Step (1)}:  For $j = 1, \dots, J$ and $r = 1, \dots, R$,  sample 
    $$\bs{\delta}_{j, r}| \pi, \bs{\tau}_{1:R}, \mb{m}_{1:R}, \bs{\delta}_{-(j,r)},  \mb{U}, \mb{V}, \mathcal{S}, \mathcal{D} \sim N( \mb{C}_{\bs{\delta}_{j, r}}\mb{g}_{\bs{\delta}_{j,r}}, \mb{C}_{\bs{\delta}_{j,r}}),$$
where $\mb{C}_{\bs{\delta}_{j,r}}  = \left( \tau_r^{-1} \bs{\Sigma}_{\bs{\delta}_j}^{-1} + \sum_{i=1}^n \bs{\Delta}_{i, j, r} \right)^{-1}$, $\mb{g}_{\bs{\delta}_{j,r}} =  \sum_{i=1}^n \bs{\chi}_{i, j, r}$,  $\bs{\Delta}_{i, j,r} =  \bs{\Xi}_{i,r} \bs{\Sigma}^{-1}_{i, (j,j)}  \bs{\Xi}_{i,r}^T$, and $\bs{\chi}_{i, j,r} = \bs{\Xi}_{i,r}  \left( \bs{\Sigma}^{-1}_{i, (j,j)} \mb{q}_{-r, i, j} + \sum_{j' \in \mathcal{J}_i, j' \neq j} \left(\bs{\Sigma}^{-1}_{i, (j',j)}  \left(\mb{q}_{-r, i. j'}  - (\bs{\delta}_{j',r}^T\mb{x}_i) \mb{v}_{r} \otimes \mb{u}_r \right) \right) \right)$.
Here, $\bs{\Sigma}_i^{-1}$ is partitioned into $J_i \times J_i$ blocks, each of size $K \times K$, with $\bs{\Sigma}_{i,(j_1, j_2)}^{-1}$ denoting  the $(j_1, j_2)$-th block.
\item \textbf{Step (2)}: For $r = 1, \dots, R$, sample 
 $$\tau_r| \pi,\bs{\tau}_{-r}, \mb{m}_{1:R},  \bs{\delta}_{1:J, 1:R},  \mb{U}, \mb{V}, \mathcal{S}, \mathcal{D} \sim \text{GIG} \left(\frac{2}{h(m_r)},  \sum_{j=1}^J \bs{\delta}_{j,r}^T\bs{\Sigma}_{\delta_j
    }^{-1}\bs{\delta}_{j,r},  -\frac{pJ}{2} + 1 \right),$$ 
    where $h(1) = h_1,  h(0) = h_0$, and $\text{GIG}(a,b,c)$ denotes
    the generalized inverse Gaussian (GIG) distribution with parameters $a>0, b > 0,$ and $c \in \mathbb{R}$.   For computational efficiency, we  approximate sampling from  $\text{GIG}(a, b, c)$  using its mode: $(c - 1 + \sqrt{(c - 1)^2 + ab}) / a$.


 \item \textbf{Step (3)}: For $r = 1, \dots, R$,  sample  
      $$m_r| \pi, \bs{\tau}_{1:R}, \mb{m}_{-r},  \bs{\delta}_{1:J, 1:R},   \mb{U}, \mb{V}, \mathcal{S},  \mathcal{D}   \sim \text{Bernoulli} \left( \tilde{p}\right),$$
where  $\tilde{p} = \pi f_{\text{HL}, h_1}(\tau_r)/ (\pi f_{\text{HL}, h_1}(\tau_r) + (1-\pi) f_{\text{HL}, h_0}(\tau_r))$. 
\item \textbf{Step (4)}: Sample 
$$\pi| \bs{\tau}_{1:R}, \mb{m}_{1:R},  \bs{\delta}_{1:J, 1:R}, \mb{U}, \mb{V}, \mathcal{S}, \mathcal{D} \sim \text{Beta}(\tilde{a}, \tilde{b}),$$
where $\tilde{a} = a_{\delta} + \sum_{r=1}^R m_r$ and $\tilde{b} = b_{\delta} + R - \sum_{r=1}^R m_r$.
\end{itemize}

To perform rank selection with shrinkage priors,  a common approach is to specify a moderately large  $R$ (exceeding the true rank) and select the rank  by thresholding component-specific parameters ($\tau_r$ in our case),  followed by restricting inference to the selected components or refitting the model with the selected rank. Instead, we implement a ``warm-start'' strategy that dynamically adds or removes ranks during sampling, which improves sampling efficiency and avoids repeated model refitting. This approach is detailed in \Cref{subsec:setup}.

\section{Simulation} \label{sec:simulation}
We refer to the proposed method in \Cref{sec:method} as Bayesian Mixed-Effects Functional Model (\textbf{BMEF}), which jointly addresses three challenges  (i) multilevel data structure, (ii) two-way functional data, and (iii) subject-level covariates.   Since no existing method simultaneously addresses all three challenges,  we benchmark our method against representative alternatives that address two of these challenges, either directly or with minor adaptations.  Our comparison includes three existing methods and two variants of our proposal:

\begin{itemize}
\item \textbf{MHPCA}:  Multilevel Hybrid Principal Components Analysis  \citep{campos2022multilevel}, which addresses (i) and can be adapted to  (ii) (but does not address (iii)); 
\item \textbf{BALMF}: Bayesian Analysis of Longitudinal and Multidimensional Functional Data \citep{shamshoian2022bayesian}, which addresses (ii) and (iii) (but not (i));
\item \textbf{FLFOSR}: Fast Longitudinal Function-on-Scalar Regression \citep{sun2024ultra}, which addresses (i) and (iii) (but not (ii)). 
\item \textbf{BMEF-1}: The proposed  Bayesian Mixed-Effects Functional Model with heterogeneous variances (see \Cref{sec:method}), which addresses (i), (ii), and (iii). 
\item  \textbf{BMEF-2}: The proposed Bayesian Mixed-Effects Functional Model  with homogeneous variance (see \Cref{sec:method} and  Supplementary Material A),  which addresses (i), (ii), and ~(iii). 
\end{itemize}

\subsection{Set up} \label{subsec:setup}
We generated data from model \eqref{eq:2d:func:model} under the basis expansion in \eqref{eq:basis:approx} and the CP decomposition in \eqref{eq:cp:decomposition}.  The simulation setup includes $J = 3$ experimental conditions, with functional responses evaluated on a $T \times F = 50 \times 50$ grid evenly spaced over the $[0,1]^2$ domain. Let $\mb{x}_i = (x_{i,1}, \dots, x_{i,p})^T$ denote the covariate vector for subject $i$, assumed to be i.i.d. realizations of a random vector $\mb{X} = (X_{1}, \dots, X_{p})^T$. We considered three scenarios with increasing complexity in the fixed effects structure, where $\mathcal{A}_j(\mb{x}_i)$ depends on
\begin{itemize}
    \item $S_1$: Condition only, with $X_{1} = 1$.  
    \item $S_2$: Condition and one feature, with $X_{1} = 1$ and $X_{2} \sim \text{Unif}(-3,3)$. 
    \item $S_3$: Condition and two features,  with $X_{1} = 1$, $X_{2} \sim \text{Unif}(-3,3)$, and $X_{3} \sim \text{Bernoulli}(0.5)$.   
\end{itemize}
For each scenario $S \in \{S_1, S_2, S_3\}$, we considered CP ranks $R \in \{2,3,4\}$.   For each $S \times R$ combination, we evaluated model performance under two variance settings 
\begin{itemize}
    \item $H_1$: Homogeneous variance, with $\sigma_{\omega_i}^2 = 0.4^2$, for $i = 1, \dots n$ 
    \item $H_2$:  Heterogeneous variance, 
 with $\sigma_{\omega_i}^2  \sim \text{Unif}(0.2^2, 1)$ independently for  $i = 1, \dots, n.$ 
\end{itemize}
We fixed $\sigma_{\gamma}^2 = 0.4^2$ and $\sigma_{\epsilon}^2 = 0.1^2$,  and generated $\bs{\gamma}_i$, $\bs{\omega}_{i,j}$, and $\epsilon_{i,j}$ based on \eqref{eq:prior:effects}. We used natural cubic  B-splines with evenly spaced interior knots as the marginal bases $\bs{\phi}(t)$ and $\bs{\psi}(f)$, setting dimensions $K_T = K_F = 6$.  The factor matrices $\mb{U}$ and $\mb{V}$ were sampled uniformly from the Stiefel manifold  $\mb{V} \in \mathcal{V}_R(\mathbb{R}^{6})$, using  the method of \cite{stewart1980efficient} implemented in the R package \texttt{pracma} \citep{pracma}. For  $j = 1,\dots,J$ and $r = 1,\dots,R$, each  element of $\bs{\delta}_{j,r}$ was independently drawn from the mixture: $0.5 \text{Unif}(-1, -0.5) + 0.5 \text{Unif}(0.5, 1)$.

\textbf{MHPCA} of \citet{campos2022multilevel} was originally developed for multilevel data with one discrete and one functional dimension but it extends naturally to two-way functional data. Under this extension, the model takes the form  
\begin{align} \label{eq:mhpca}
    Y_{i,j}(t,f) =    \mathcal{A}_{j}(t,f) +       \mathcal{B}_{i}(t,f) +   \mathcal{C}_{i,j}(t,f) +  \mathcal{E}_{i,j}(t,f),
    \end{align}
where $\mathcal{A}_{j}$ is estimated by smoothing the average of the functional observations $\mathcal{Y}_{i,j}(\mb{t},\mb{f})$ across subjects, and $\mathcal{B}_{i}$ and $\mathcal{C}_{i,j}$ are modeled using eigenfunctions of the subject- and subject-by-condition-level covariances, respectively. The noise term $\mathcal{E}_{i,j}(t,f)$ is assumed to be  independent of $t$ and $f$.   Estimation is performed under a weak separability assumption, where the eigenfunctions of the two-way covariance are constructed from those of the marginal covariances smoothed using tensor product bases. Unlike the original implementation of \textbf{MHPCA}, which uses discrete-continuous bases, we used  continuous marginal bases in both dimensions to accommodate two-way functional covariances, with 6 basis functions per dimension.  
The implementation was adapted from the R package \texttt{mhpca} \citep{mhpca}.


For \textbf{BALMF}, we adopted the implementation from the R package \texttt{LFBayes}  \citep{LFBayes}. Since \textbf{BALMF}  does not support multilevel data,  we fit separate models for each condition  $j = 1, \dots J$,  yielding the following model:
\begin{align*} 
    Y_{i,j}(\mb{x}_i)(t,f) =      \mathcal{A}_{j}(\mb{x}_i)(t,f) +       \mathcal{C}_{i,j}(t,f) + \mathcal{E}_{i,j}(t,f), \ i \in \{i', j \in \mathcal{J}_{i'}\}.
    \end{align*}
\textbf{BALMF} shares several features with our proposed  \textbf{BMEF}, including the use of  tensor-product bases to approximate $\mathcal{A}_{j}(\mb{x})$ and $\mathcal{C}_{i,j}$, and a covariate-dependent factor model for $\mathcal{A}_{j}(\mb{x})$, however, its factor model relies on a Tucker-type decomposition,  which is unidentifiable and less parsimonious than the CP decomposition used in \textbf{BMEF}.   We applied \textbf{BALMF} with natural cubic B-splines (6 knots) as marginal bases  and set the factor ranks to $4 \times 4$, resulting in greater model complexity than our highest CP-rank setting ($R = 4$). \textbf{BALMF} incorporates a rank-regularization prior that avoids manual rank selection \citep{shamshoian2022bayesian}. We ran the MCMC sampler for 1000 burn-in iterations followed by 400 posterior draws.

\textbf{FLFOSR} was originally developed for one-way functional data \citep{sun2024ultra}. To apply it to two-way functional observations, we fit separate models for each frequency grid point $f \in \{f_1, \dots, f_F\}$, resulting in the following model:
\begin{align*}
    Y_{i,j, f}(\mb{x}_i)(t) =      \mathcal{A}_{j, f}(\mb{x}_i)(t) +   \mathcal{B}_{i,j, f}(t) +     \mathcal{C}_{i,j, f}(t)  +   \mathcal{E}_{i,j, f}(t),
\end{align*}
where $\mathcal{A}_{j, f}(\mb{x}_i)$, $\mathcal{B}_{i,j, f}$, and  $\mathcal{C}_{i,j, f}$ are one-way functions, and $ \mathcal{E}_{i,j, f}(t)$ represents random noise independent of $t$.  We used the R package \texttt{FLFOSR}  \citep{flfosr} to fit the model, approximating $\mathcal{A}_{i,j,f}(\mb{x}_i)(t)$, $\mathcal{B}_{i,j,f}(t)$, and $\mathcal{C}_{i,j,f}(t)$ using cubic P-splines of dimension 6.  The original implementation of \textbf{FLFOSR}  assumes a common fixed effect function shared  across conditions.  To ensure a fair comparison with \textbf{BMEF}, we modified the input design matrix to include condition-specific indicators and their interactions with covariates, enabling the fixed effect function $\mathcal{A}_{j,f}$ to vary across both conditions and covariates.  We ran the method with 1000 burn-in iterations followed by 400  sampling iterations.  Posterior draws of $\mathcal{A}_{i,j}(\mb{x}_i)(t, f)$, $\mathcal{B}_{i,j}(t, f)$, and $\mathcal{C}_{i,j}(t, f)$ were obtained by concatenating the posterior samples of $\mathcal{A}_{j, f}(\mb{x}_i)$,   $\mathcal{B}_{i,j, f}$, and  $\mathcal{C}_{i,j, f}$ over $f \in \{f_1, \dots, f_F\}$.

For  \textbf{BMEF-1} and \textbf{BMEF-2}, we specified the marginal bases  $\bs{\phi}(t)$ and $\bs{\psi}(f)$  as natural cubic B-splines with dimensions $K_T = K_F = 6$ and evenly spaced interior knots. The prior parameters were set as follows: $a_{\gamma} = 3$, $b_{\gamma} = 0.5$, $a_{\omega} = 3$, and $b_{\omega} = 0.5$ in \eqref{eq:prior:var}; $\bs{\Sigma}_{\delta_j} = 5 \mb{I}p$ in \eqref{eq:prior:delta};  $a_{\delta} = 1$, $b_{\delta} = 1$, $h_1 = 1$, and $h_0 = 0.01$ in \eqref{eq:prior:ssl}.  We used uniform priors for $\mb{U}$ and $\mb{V}$. The MCMC sampler was run with 800 burn-in iterations and 400 sampling iterations. To select the CP rank, we used a ``warm-start'' strategy  that adaptively adjusts the rank during sampling to accelerate convergence. The algorithm begins with rank 1 and evaluates whether to add or remove ranks every 100 iterations.   If all $\tau_r$ exceed a threshold (set to 0.05), an additional rank is introduced. If any $\tau_r$ falls below the threshold, the corresponding rank is removed.  Once a rank is removed,  the algorithm does not consider adding new ranks in subsequent iterations. When an irrelevant rank $r'$ is introduced, our approach yields sparser estimates for $\bs{\delta}_{j,r'}$  compared to forward selection without a shrinkage prior, and the values of $\tau_r$'s provide a practical and effective guide for rank selection. 
For the competing methods (\textbf{MHPCA}, \textbf{BALMF}, \textbf{FLFOSR}), any model parameters not mentioned above were set to their default values. 

\subsection{Evaluation metrics}
We conducted 100 independent simulation runs, randomly generating $\mb{x}_i$, $\bs{\alpha}_j$, $\bs{\gamma}_i$, $\bs{\omega}_{i,j}$, and $\bs{\epsilon}_{i,j}$ in each run. The performance of the methods under comparison was evaluated in the following aspects. 

\subsubsection{Estimation of fixed effects}
We evaluated the estimation accuracy of  $\mathcal{A}_j$ using  mean squared error (MSE) defined as 
    $\text{MSE}(\mathcal{A}) =\sum_{i=1}^n \sum_{j=1}^J  ||\hat{\mathcal{A}}_j(\mb{x}_i)(\mb{t}, \mb{f}) - \mathcal{A}_j(\mb{x}_i)(\mb{t}, \mb{f}) ||_F^2/(nJ),$ 
    where $||\cdot||_F$ denotes the Frobenius norm. For Bayesian methods (\textbf{BALMF}, \textbf{FLFOSR}, \textbf{BMEF-1}, and \textbf{BMEF-2}), $\hat{\mathcal{A}}_j(\mb{x}_i)(\mb{t}, \mb{f})$ denotes the posterior mean; for \textbf{MHPCA}, $\hat{\mathcal{A}}_j(\mb{x}_i)(\mb{t}, \mb{f})$ corresponds to the  point estimate.  

In $S_2$ and $S_3$ settings, we  evaluated the accuracy of  \textbf{BALMF}, \textbf{FLFOSR}, \textbf{BMEF-1}, and \textbf{BMEF-2} in estimating the covariate effects using a contrast mean squared error (CMSE). For  $X_2$ in $S_2$ and $S_3$, and $X_3$ in $S_3$, we define 
$\text{CMSE}(X_k) = \sum_{i=1}^n \sum_{j=1}^J  || \widehat{\text{Diff}}(X_k)(\mb{t}, \mb{f})  - \text{Diff}(X_k)(\mb{t}, \mb{f})||_F^2/(nJ)$, for $k \in \{2, 3\}$, where  $\text{Diff}(X_k)$  quantifies the effect of changing $X_k$ from 0 to 1, while conditioning on the other covariates being held fixed.    For $S_2$,  $\text{Diff}(X_2)(\mb{t}, \mb{f}) = \mathcal{A}_j( (X_1, 1))(\mb{t}, \mb{f}) - \mathcal{A}_j((X_1, 0))(\mb{t}, \mb{f})$; for $S_3$,  $\text{Diff}(X_2)(\mb{t}, \mb{f}) = \mathcal{A}_j((X_1, 1, X_3))(\mb{t}, \mb{f}) -\mathcal{A}_j((X_1, 0, X_3))(\mb{t}, \mb{f})$,  and $\text{Diff}(X_3) = \mathcal{A}_j((X_1, X_2, 1))(\mb{t}, \mb{f}) - \mathcal{A}_j((X_1, X_2, 0))(\mb{t}, \mb{f})$.   The  estimated effects, denoted $\widehat{\text{Diff}}(X_2)$ and  $\widehat{\text{Diff}}(X_3)$, were computed by replacing $\mathcal{A}_j$ with  $\hat{\mathcal{A}}_j$.  Due to the additive structure of covariate effects in the models under comparison,  the values of the conditioning covariates do not affect $\text{Diff}(X_k)$ and $\widehat{\text{Diff}}(X_k)$ for  $k \in \{2,3\}$.  

For \textbf{BMEF-1} and \textbf{BMEF-2}, we further evaluated their ability to correctly select the true rank $R$, using the sparsity-inducing prior on covariate coefficients introduced in \Cref{subsubsec:covariate}.   

\subsubsection{Estimation of random effects}
We used MSE to assess the accuracy of estimating subject- and subject-by-condition-level random effects.   For \textbf{MHPCA}, \textbf{FLFOSR}, \textbf{BMEF-1}, and \textbf{BMEF-2}, the MSE for subject-specific effects was computed as $\text{MSE}(\mathcal{B}) = \sum_{i=1}^n || \hat{\mathcal{B}}_i(\mb{t}, \mb{f}) - \mathcal{B}_i(\mb{t}, \mb{f}) ||_F^2/n$, and the  MSE for subject-by-condition-level  effects was computed for all methods as  $\text{MSE}(\mathcal{C}) =\sum_{i=1}^n \sum_{j\in \mathcal{J}_i} || \hat{\mathcal{C}}_{i,j}(\mb{t}, \mb{f}) - \mathcal{C}_{i,j}(\mb{t}, \mb{f})||_F^2/(nJ')$.  For \textbf{FLFOSR}, \textbf{BMEF-1}, and \textbf{BMEF-2},  the estimates $\hat{\mathcal{B}}_i(\mb{t}, \mb{f})$ and  $\hat{\mathcal{C}}_{i,j}(\mb{t}, \mb{f})$ correspond to the respective posterior means, while for \textbf{MHPCA}, they are point estimates.  \textbf{BALMF} does not include $\mathcal{B}_i$ in its model, and its $\hat{\mathcal{C}}_{i,j}(\mb{t}, \mb{f})$ was computed from the posterior mean.

We report results without missing conditions (i.e., $\mathcal{J}_i = \{1, \dots, J\}$ for $i = 1, \dots, n$) in \Cref{subsec:results} below. Additional simulation results for scenarios with missing conditions are reported in Supplementary Material~E.

\subsection{Results}  \label{subsec:results}

\begin{figure}[htp]
\includegraphics[scale=0.55]{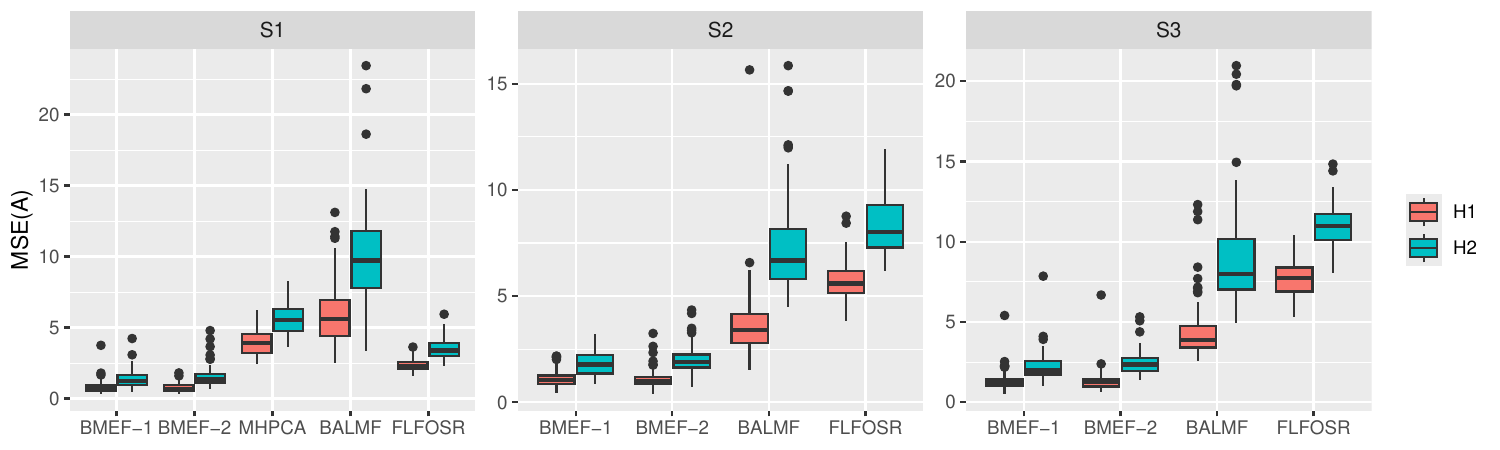}
\caption{$\text{MSE}(\mathcal{A})$ from 100 simulation runs for rank $R = 4$, under  $\{S_1, S_2, S_3\}$ and $\{H_1, H_2\}$ combinations.}
\label{fig:mseA}
\end{figure}

\begin{figure}[htp]
\includegraphics[scale=0.55]{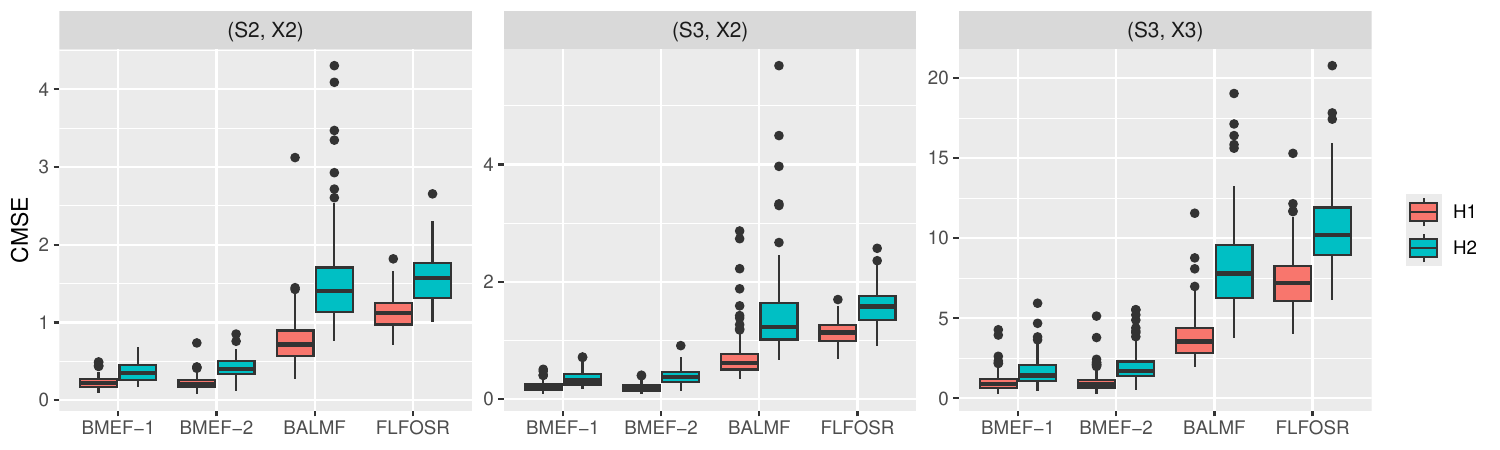}
\caption{$\text{CMSE}$ from 100 simulation runs for rank $R = 4$, displaying $\text{CMSE}(X_2)$ for $S_2$ and $S_3$, and $\text{CMSE}(X_3)$ for $S_3$, across $\{H_1, H_2\}$ settings.}
\label{fig:cmseA}
\end{figure}

We  present results  for the most complex setting with
rank $R = 4$. Findings from other settings follow similar trends and are summarized in this section, with details provided in Supplementary Material E.   Results for \textbf{MHPCA}  are only reported for $S_1$ settings for the fixed effect, as this method does not account for covariate effects and performs substantially worse in covariate-dependent settings $S_2$ and $S_3$. Although \textbf{MHPCA} includes $\mathcal{B}_i$ and $\mathcal{C}_{i,j}$, their estimation is highly unstable and therefore not reported here. \Cref{fig:mseA} presents  $\text{MSE}(\mathcal{A})$ from 100 simulation runs.  Across all covariate settings ($S_1$, $S_2$, and $S_3$),  \textbf{BMEF-1} and \textbf{BMEF-2} consistently achieve the lowest  $\text{MSE}(\mathcal{A})$ under both variance configurations $H_1$ and $H_2$. Notably, \textbf{BMEF-1} performs comparably to \textbf{BMEF-2} under $H_1$, and outperforms \textbf{BMEF-2} under $H_2$, demonstrating the flexibility of \textbf{BMEF-1} in accommodating heterogeneous variances across subjects.  \Cref{fig:cmseA} presents CMSE  for estimating the covariate effects based on 100 simulations runs. Consistent with the $\text{MSE}(\mathcal{A})$ findings, 
\textbf{BMEF-1} and  \textbf{BMEF-2} outperform all competing methods across settings. Under $H_2$,   \textbf{BMEF-1} achieves lower CMSEs compared to \textbf{BMEF-~2}.  The poor performance of \textbf{BALMF} and \textbf{FLFOSR} results from
their inability to share information across dimensions and conditions. \textbf{FLFOSR} fits separate models for each frequency, ignoring the smoothness across the frequency domain.  \textbf{BALMF} fits each condition independently, without incorporating subject-level random effects shared across conditions.  In  $S_1$ settings, although \textbf{MHPCA} shares the same model \eqref{eq:mhpca} as \textbf{BMEF-1} and \textbf{BMEF-2}, its reliance on the weak separability assumption on the functonal covariances and naive use of smoothing splines to estimate the fixed effects without accounting for the random effects \citep{campos2022multilevel} can lead to model misspecification and inferior estimation. 

\begin{figure}[htp]
\includegraphics[scale=0.55]{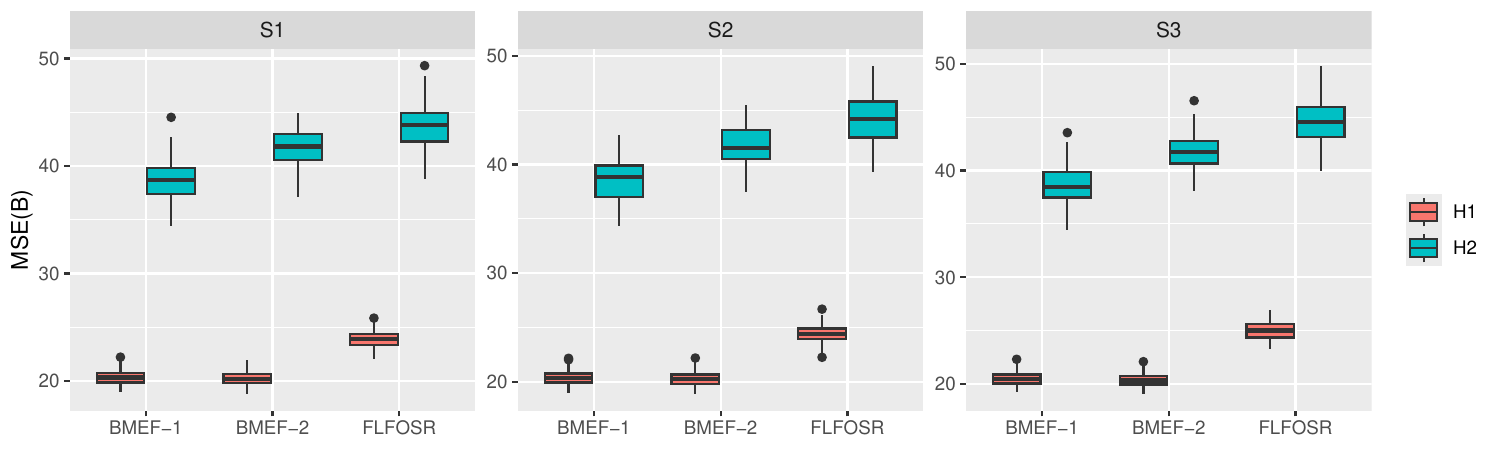}
\caption{$\text{MSE}(\mathcal{B})$ from 100 simulation runs for rank $R = 4$, under  $\{S_1, S_2, S_3\}$ and $\{H_1, H_2\}$ combinations.}
\label{fig:mseB}
\end{figure}

\begin{figure}[htp]
\includegraphics[scale=0.55]{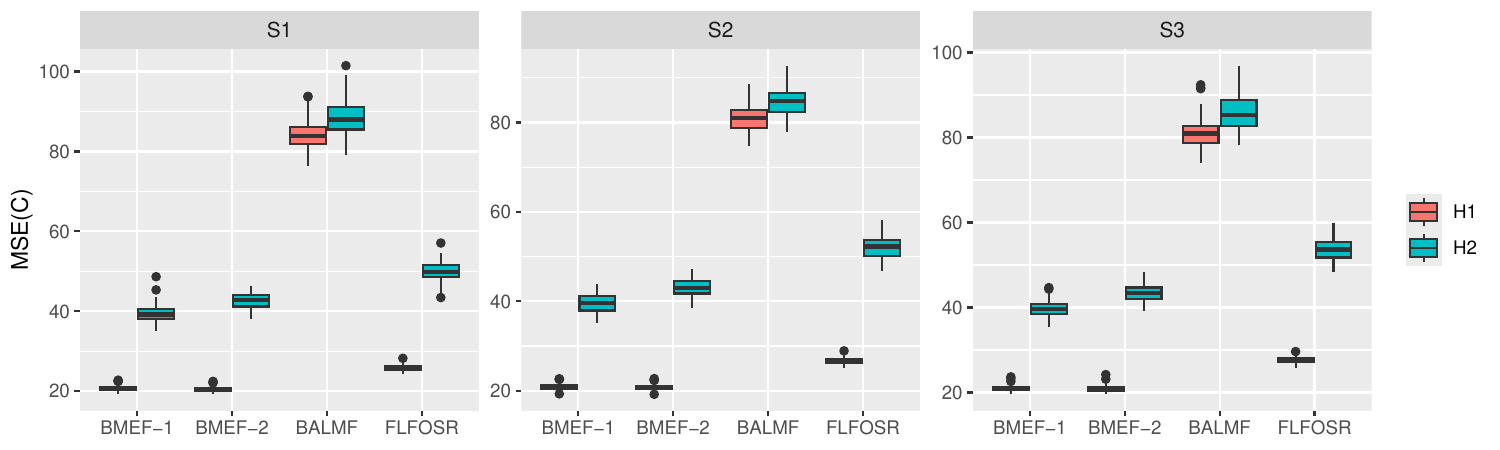}
\caption{$\text{MSE}(\mathcal{C})$ from 100 simulation runs for rank $R = 4$,  under  $\{S_1, S_2, S_3\}$ and $\{H_1, H_2\}$ combinations.}
\label{fig:mseC}
\end{figure}

\Cref{fig:mseB} and \Cref{fig:mseC} present the results for $\text{MSE}(\mathcal{B})$ and  $\text{MSE}(\mathcal{C})$, respectively.  \textbf{BALMF} is excluded from \Cref{fig:mseB} as it does not explicitly model $\mathcal{B}_i$,  instead absorbing its effect into the estimated $\mathcal{C}_{i,j}$. Both figures show that \textbf{BMEF-1} and \textbf{BMEF-2} outperform the competing methods, with \textbf{BMEF-1} showing better performance compared to \textbf{BMEF-2} under $H_2$.

Results in  Supplementary Material E show that for ranks $R=2$ and $R=3$, all methods show similar performance patterns to those for $R = 4$, leading to the same  conclusion: \textbf{BMEF-1} and \textbf{BMEF-2} consistently achieve the best performance, with \textbf{BMEF-1} showing greater flexibility under $H_2$ settings.  As the rank increases from 2 to 4, $\text{MSE}(\mathcal{A})$ increases for all methods, with a more substantial increase for \textbf{BALMF}  in $S_1$ settings and for all competitors in $S_2$ and $S_3$ settings. Across all ranks,  \textbf{FLFOSR} is especially sensitive to covariate complexity, showing degraded performance from  $S_1$ to $S_3$, likely due to the  increased feature dimensionality from covariate-by-condition interactions. In contrast, \textbf{BMEF-1} and \textbf{BMEF-2} maintain stable performance across settings, showing only a slight increase in $\text{MSE}(\mathcal{A})$ and $\text{MSE}(\mathcal{C})$  and comparable $\text{MSE}(\mathcal{B})$  with the inclusion of more  covariates.  

Lastly, we assessed the rank selection accuracy of \textbf{BMEF-1} and \textbf{BMEF-2} across all settings by computing the proportion of simulation runs (out of 100) in which the selected rank matched the true rank. Both methods correctly selected the true rank in 100\% of the runs for $S_2$ and $S_3$ settings, and in 99\% of the runs for $S_1$ settings, demonstrating the effectiveness of the proposed rank selection procedure. Detailed  are provided in Supplementary Material E.

\section{Application}  \label{sec:application}
Understanding the neural signatures of alcohol-related cognitive dysfunction remains an important challenge in clinical neuroscience, particularly as alcoholism is associated with altered frontal lobe activity and impaired working memory. We applied our method to an EEG dataset collected to study how alcoholism affects cognitive processing in response to visual stimuli.
 The dataset includes recordings from 122 subjects (77
alcoholic, 45 nonalcoholic), collected  
by the Neurodynamics Laboratory at the State University of New York Health Center and is  available from the UCI Machine Learning Repository \citep{eeg_database_121}.  
Each subject completed 120 trials,  with each trial presenting one of three visual stimulus types: a single image ($C_1$), a pair of matched images ($C_2$), or a pair of non-matched images ($C_3$) \citep{zhang1995event}.  After excluding artifact-contaminated trials, an average of 91 usable trials per subject was retained.  
Time-frequency representations of EEG signals were computed using the \texttt{spectro} function from the R package \texttt{seewave}  \citep{seewave}, applying a short-time Fourier transform and yielding power spectra on a grid of  $T = 21$ time points and $F = 25$ frequency points. Power was expressed in decibels (dB), and we applied 1/f correction \citep{gyurkovics2021impact}  by subtracting the mean dB across time and subjects to reduce spectral imbalance across frequency. For each subject, the time-frequency representations were averaged across trials per condition. To focus on frontal brain activity implicated in cognitive control and working memory \citep{cavanagh2014frontal}, we analyzed data from the AFz electrode, where preliminary analyses showed representative spectral patterns across all regions.  


We investigated how alcoholism (coded as a binary indicator) is associated with multi-condition EEG time-frequency responses and compared three \textbf{BMEF} model specifications of decreasing complexity. All models include an intercept and the alcoholism indicator as covariates, but differ in their random effects specification at the subject and subject-by-condition level:
\begin{itemize}
  \item  \textbf{BMEF-ABC}:  Includes both subject-level $(\mathcal{B}_i)$ and subject-by-condition-level $(\mathcal{C}_{i,j})$ random effects
  $Y_{i,j}(\mb{x})(t,f) = \mathcal{A}_j(\mb{x})(t,f) +  \mathcal{B}_i(t,f) + \mathcal{C}_{i,j}(t,f) + \mathcal{E}_{i,j}(t,f)$
    \item  \textbf{BMEF-AB}:   Includes only subject-level random effects 
    $Y_{i,j}(\mb{x})(t,f) = \mathcal{A}_j(\mb{x})(t,f) +  \mathcal{B}_i(t,f) + \mathcal{E}_{i,j}(t,f)$
      \item  \textbf{BMEF-A}:  Includes no random effects $Y_{i,j}(\mb{x})(t,f) = \mathcal{A}_j(\mb{x})(t,f) + \mathcal{E}_{i,j}(t,f)$
\end{itemize}
For these models, we conducted inference using the \textbf{BMEF-1} and \textbf{BMEF-2} frameworks introduced in \Cref{sec:simulation}, which assume heterogeneous and homogeneous variances, respectively.  For both \textbf{BMEF-1} and \textbf{BMEF-2}, for rank selection, three threshold values for $\tau_r$ in \eqref{eq:prior:delta}, denoted by $\lambda_{\tau} \in \{0.01, 0.05, 0.1\}$, were considered.  The ``warm-start'' procedure introduced in \Cref{subsec:setup} was adopted.  Natural cubic B-splines with evenly spaced interior knots were used as marginal basis functions, with dimensions $K_T = 6$ and $K_F = 10$. These dimensions were found to be sufficient, as increasing their values did not improve the estimation of the underlying functional patterns. Other implementation details remained the same as those described in \Cref{subsec:setup}.  The performance of each model was evaluated using  Watanabe–Akaike Information Criterion (WAIC) \citep{vehtari2017practical} defined as 
\begin{align}\label{eq:waic}
    \text{WAIC} = -2 \left( \sum_{i=1}^n \log \left( \frac{1}{S} \sum_{s=1}^S p \left(\{\mb{y}_{i,j}\}_{j \in \mathcal{J}_i} \mid \bs{\Theta}_s \right) \right) - \sum_{i=1}^n \text{Var}\left( \log  \left(p \left(\{\mb{y}_{i,j}\}_{j \in \mathcal{J}_i}|\bs{\Theta}_s \right) \right) \right) \right),
\end{align}
where $\bs{\Theta}_s$ denotes the set of  parameters for the corresponding model specification at the $s$-th posterior draw, and $S$ is the total number of posterior samples. 
We repeated the experiment 100 times,  each time fitting all models using data from a randomly selected 80\% subset of the subjects.

\begin{table}
\centering
\caption{WAIC summary (mean (SD)) for \textbf{BMEF-1} and \textbf{BMEF-2} across thresholds ($\lambda_{\tau} \in \{0.01, 0.05, 0.1\}$) and the three model specifications, based on 100 experimental runs. Reported values are scaled by a factor of $10^{-2}$.}
\begin{tabular}{lcccccc}
\toprule
 & \multicolumn{3}{c}{\textbf{BMEF-1}} & \multicolumn{3}{c}{\textbf{BMEF-2}} \\
\cmidrule(lr){2-4} \cmidrule(lr){5-7}
 & $\lambda_{\tau} = 0.01$ & $\lambda_{\tau} = 0.05$ & $\lambda_{\tau} = 0.1$ & $\lambda_{\tau} = 0.01$ & $\lambda_{\tau} = 0.05$ & $\lambda_{\tau} = 0.1$\\
\midrule
\textbf{BMEF-ABC} & 2672.4 (19.5) & 2672.6 (19.3) & 2672.8 (19.6) & 2665.3 (19.7) & 2665.2 (19.5) & 2665.5 (19.7)  \\
\textbf{BMEF-AB}  & 3842.7 (46.3) & 3843.0 (46.1) & 3842.9 (45.9) & 3842.7 (46.3) & 3843.0 (46.1) & 3842.9 (45.9) \\
\textbf{BMEF-A}   &  5765.8 (65.4) & 5765.7 (65.3) & 5765.7 (65.4) & 5765.8 (65.4) & 5765.7 (65.3) & 5765.7 (65.4) \\
\bottomrule
\end{tabular}
\label{eeg:waic}
\end{table}

\Cref{eeg:waic} summarizes WAIC values across model specifications and threshold levels  $\lambda_{\tau} \in \{0.01, 0.05, 0.1\}$.   \textbf{BMEF-ABC}  consistently outperforms  \textbf{BMEF-AB} and \textbf{BMEF-A}, suggesting the importance of incorporating both subject-level and subject-by-condition-level random effects $\mathcal{B}_i$ and $\mathcal{C}_{i,j}$. The improvement from including $\mathcal{B}_i$ (via \textbf{BMEF-AB}) over \textbf{BMEF-A} suggests substantial between-subject variability for all visual stimuli, and the additional gain from including $\mathcal{C}_{i,j}$ (via  \textbf{BMEF-ABC}) suggests strong  individualized responses to distinct stimulus conditions, with both sources of variability unrelated to alcoholism. Under the \textbf{BMEF-ABC} specification, \textbf{BMEF-2} yields  improved performance compared to \textbf{BMEF-1}  with respect to WAIC, suggesting that the heterogeneous variance assumption in \textbf{BMEF-1} may introduce unnecessary complexity and lead to overfitting. WAIC values are stable across different threshold ($\lambda_{\tau}$) values. For \textbf{BMEF-1} and \textbf{BMEF-2}, setting $\lambda_{\tau} = 0.01$ typically selected rank $R = 4$, $\lambda_{\tau} = 0.05$ yielded variable selections of $R = 3$ or 4, and  $\lambda_{\tau} = 0.1$ selected $R = 2$ across runs.  When $R = 3$ or 4 were selected, we consistently observed two dominant ``base'' patterns that receive much higher weights compared to the rest.  Notably, these two patterns closely matched those from the $R = 2$ model, suggesting models with $R>2$ introduced redundant or weakly contributing components. Based on these observations, we performed the full analysis on all 122 subjects using  using \textbf{BMEF-2} with $\lambda_{\tau} = 0.1$, under the \textbf{BMEF-ABC} specification, and report the findings below.
\begin{figure}
\centering  
\includegraphics[scale=0.19]{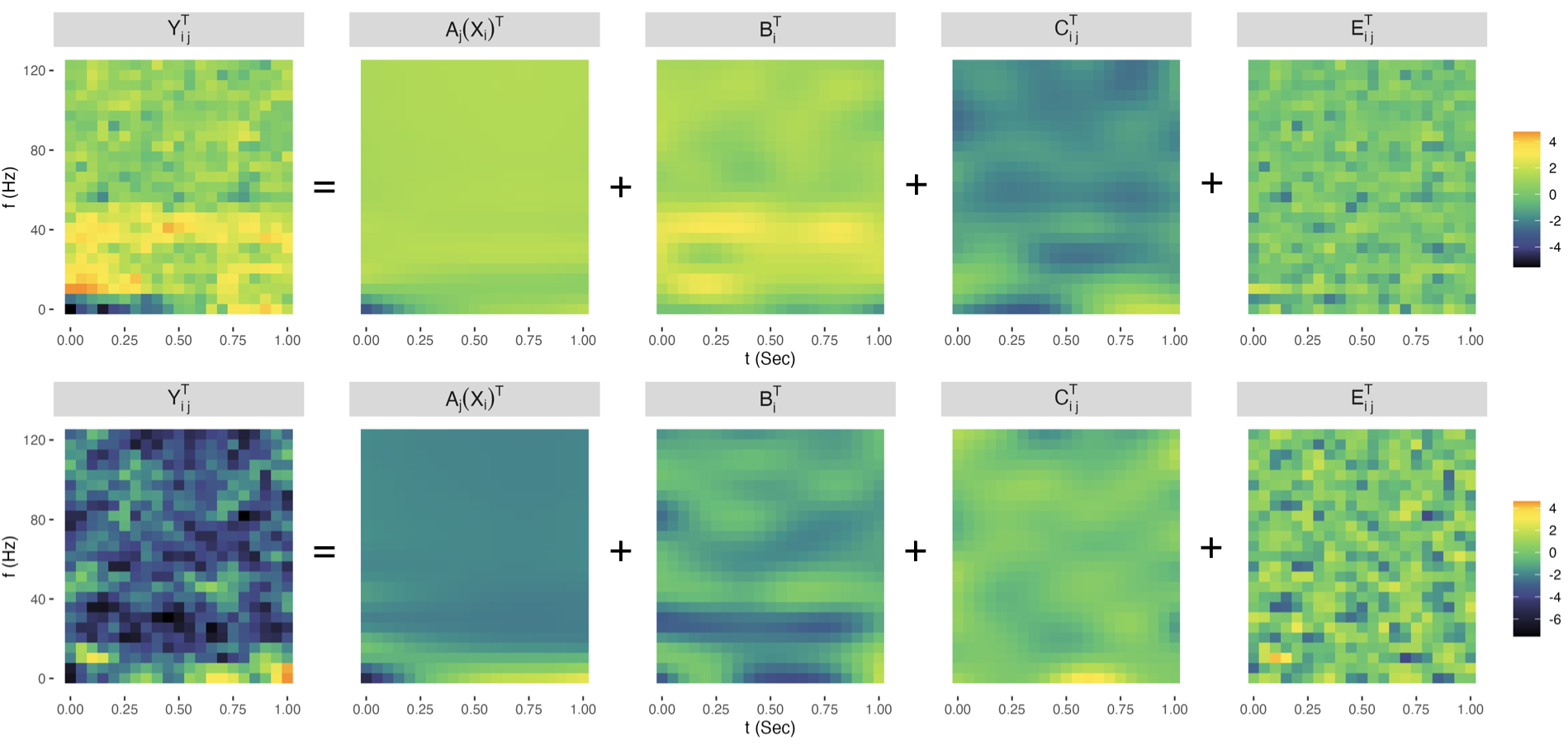}  
\caption{Decompositions of the time-frequency response $\mb{Y}_{i,j} = \mb{A}_j(\mb{x}_i) + \mb{B}_i + \mb{C}_{i,j} + \mb{E}_{i,j} $, where  $\mb{A}_j(x_i)$, $\mb{B}_i$, $\mb{C}_{i,j}$, and $\mb{E}_{i,j}$ are evaluations of $\mathcal{A}_j(x_i)$, $\mathcal{B}_i$, $\mathcal{C}_{i,j}$, and  $\mathcal{E}_{i,j}$ on a $\mb{t}\times \mb{f}$ grid of dimensions $21 \times 25$.  Results are shown for one representative alcoholic subject under condition $C_1$ (presented with ``single image'') (top row) and one representative control subject under condition  $C_2$ (presented with ``matched images'') (bottom row).} 
\label{eeg:decomp}
\end{figure}

\Cref{eeg:decomp} illustrates the model-based decomposition of the observed time-frequency representation $\mb{Y}_{i,j} \in \mathbb{R}^{21 \times 25}$  for two representative subject-condition pairs:  an alcoholic subject (Subject 1) under condition $C_1$, and a control subject (Subject 2) under condition $C_2$.   The evaluations of the estimated $\mathcal{A}_j(\mb{x}_i)$, $\mathcal{B}_i$, $\mathcal{C}_{i,j}$ and  $\mathcal{E}_{i,j}$ over the observed  time-frequency grid  are denoted by $\mb{A}_j(\mb{x}_i)$, $\mb{B}_i$, $\mb{C}_{i,j}$ and  $\mb{E}_{i,j}$, respectively.  From \Cref{eeg:decomp}, the fixed effect $\mb{A}_j(\mb{x}_i)$ differs notably between Subject 1 and 2, reflecting variation attributable to alcoholism and visual stimulus type.  The subject-level random effect $\mb{B}_i$ shows distinct time-frequency patterns between  Subjects 1 and 2, particularly in the low to mid-frequency range:  Subject~1 shows elevated activity around 10–40 Hz, and Subject 2 shows reduced activity around 10–30 Hz. These patterns can be observed in their corresponding  raw responses $\mb{Y}_{i,j}$.  The subject-by-condition random effect  $\mb{C}_{i,j}$ appears smooth and captures the remaining functional structure. The residual term $\mb{E}_{i,j}$ appears unstructured, supporting the Gaussian noise assumption imposed by our model.

\begin{figure}
\centering
\includegraphics[scale=0.47]{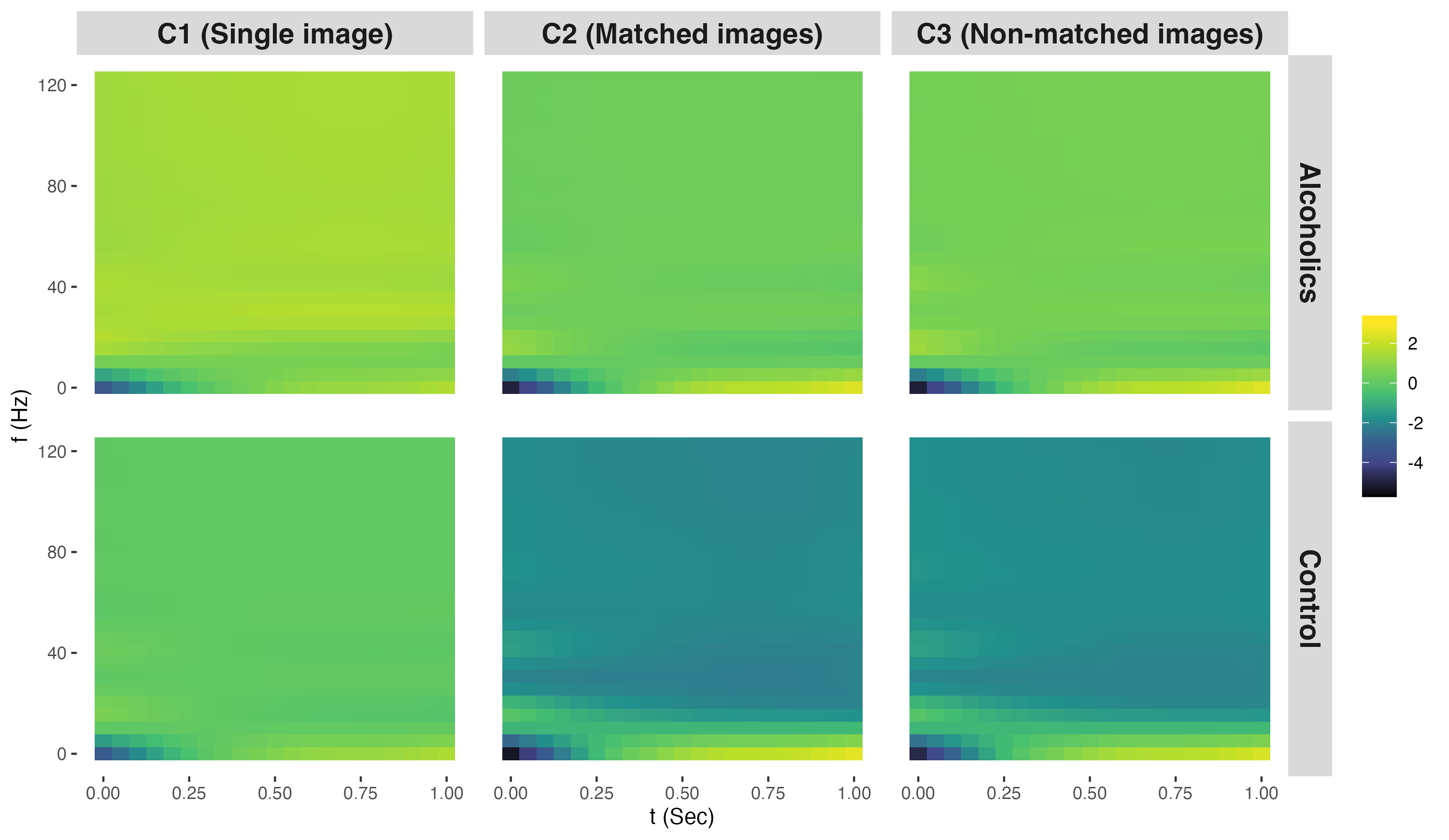}  
\caption{Posterior means of the stimulus condition- and group-specific fixed effect $\mathcal{A}_j(\mb{x}_i)$  evaluated on a $\mb{t} \times \mb{f}$ grid, shown for combinations of conditions ($C_1$, $C_2$, $C_3$) and groups (Alcoholics vs. Control), highlighting differences across conditions and groups. } 
\label{fig:eeg:fixed}
\end{figure}

 \Cref{fig:eeg:fixed} displays the posterior means  of the fixed effect $\mathcal{A}_j(\mb{x}_i)$ across combinations of experimental conditions ($C_1$, $C_2$, and $C_3$) (columns) and alcoholism status (rows).   Alcoholic subjects show higher estimated fixed effect values across conditions, particularly in the higher frequency bands.  Additionally, the fixed effect pattern for $C_2$ and $C_3$ are more similar to each other than to $C_1$,  consistent with the experimental design: $C_2$ and $C_3$ involve paired visual inputs, while $C_1$ consists of a single image input.

\begin{figure}
\hspace{-0.6cm}
    \includegraphics[scale=0.37]{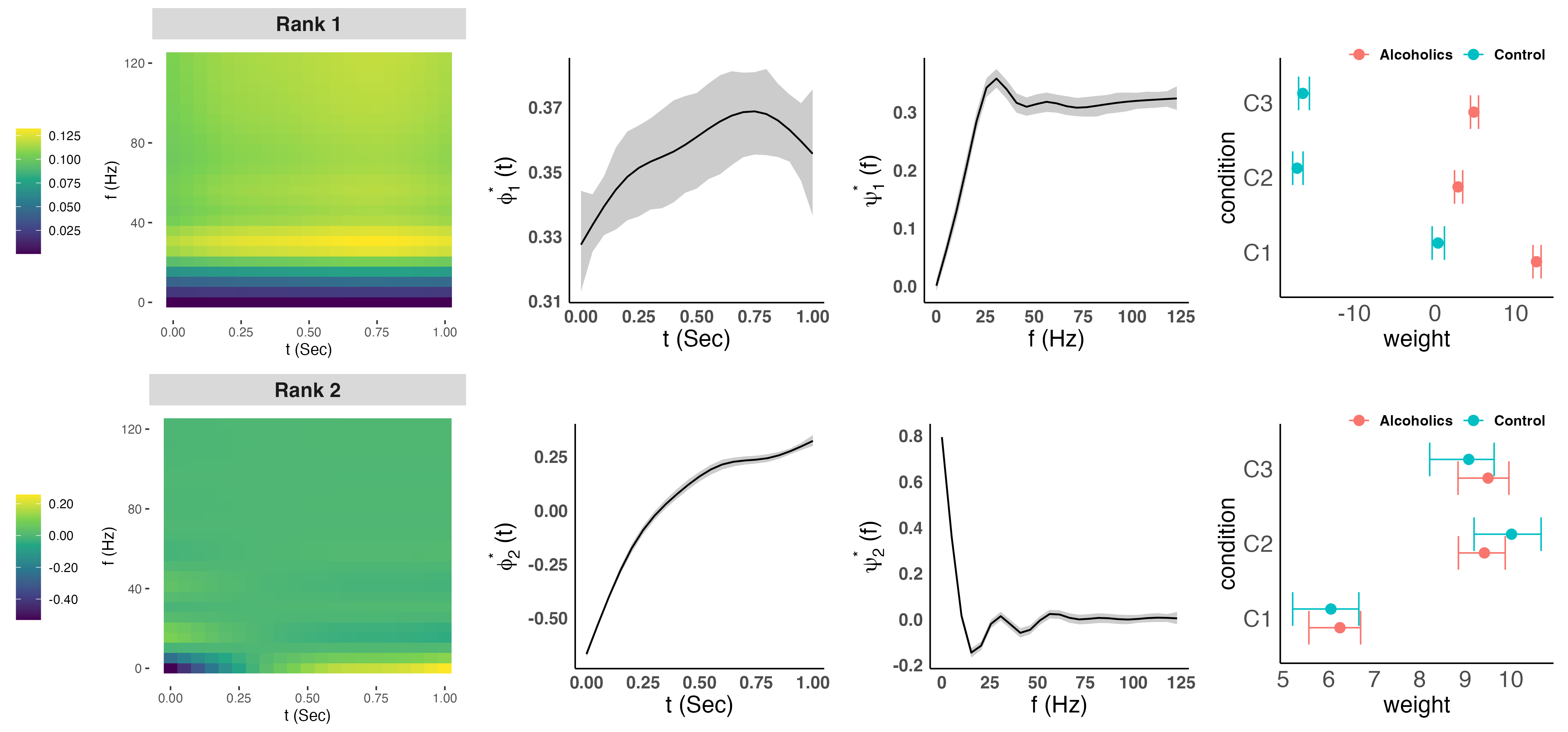}
    \caption{Visualization of rank-specific patterns and  associated weights. Each row corresponds to a rank component ($r = 1$ and $r = 2$), and columns (from left to right) display: (1) posterior mean of the ``base'' pattern $\phi_r^{\ast}(t)\psi_r^{\ast}(f)$; (2) posterior mean and 95\% credible interval of the time  ``principal''  function $\phi_r^{\ast}(t)$; (3) posterior mean and 95\% credible interval of the frequency ``principal'' function $\psi_r^{\ast}(f)$;  and (4) 95\% credible intervals of the weights $\lambda_{j,r}(\mb{x}) = \bs{\delta}_{j,r}^T\mb{x}$, shown across combinations of conditions $(C_1, C_2, C_3)$ and groups (Alcoholics vs. Control).}
        \label{fig:eeg:pattern}
\end{figure}

\Cref{fig:eeg:pattern} presents the estimated rank-specific ``base'' patterns $\phi_r^{\ast}(t)\psi_r^{\ast}(f)$ that form the fixed effect $\mathcal{A}_j(\mb{x}_i)$,  along with posterior summaries of the
associated  ``principal'' functions  $\phi_r^{\ast}(t)$ and $\psi_r^{\ast}(f)$ and the covariate-modulated weights $\lambda_{j,r}(\mb{x}) = \bs{\delta}_{j,r}^T\mb{x}$. In \Cref{fig:eeg:pattern}, the Rank~1  ``base'' pattern primarily captures gamma-band (>30 Hz) activity with a modest increase over time, as reflected by the rising trend in $\phi_{1}(t)$ and elevated  $\psi_{1}(f)$ values at high frequencies. The 95\% credible intervals for the corresponding weights $\lambda_{j,1}(\mb{x})$ (the right-most column) indicate a significant difference between alcoholic and control groups under all conditions $(C_1, C_2, C_3)$, suggesting that alcoholic subjects exhibit stronger gamma-band activity in response to the presentation of visual stimulus, compared to controls. This finding aligns with prior analyses of the same dataset under condition $C_1$, which reported similar group difference in gamma-band activity \citep{bavkar2019rapid, qazi2021electroencephalogram}.  Our analysis extends these results by capturing both spectral and temporal dynamics and showing that the group gamma-band difference generalizes across all conditions, while also accounting for subject-specific neural responses unexplained by covariates. Moreover, we found that this group difference is more pronounced under conditions $C_2$ and $C_3$,  which involve more complex visual inputs (two images), compared to  $C_1$ (a single image).  The Rank 2 ``base'' pattern mainly reflects low-frequency activities (delta, theta and alpha bands; <12 Hz), with an increasing temporal trend.  This trend aligns with established functional regulatory roles, where low-frequency oscillations  are associated with functional inhibition \citep{bonnefond2013role}, which tends to  become more pronounced during post-stimulus transition towards the resting state.  In addition, the weights $\lambda_{j,2}(\mb{x})$ differ significantly between condition $C_1$ and conditions $C_2$ and $C_3$, but not between alcoholic and control groups (\Cref{fig:eeg:pattern}, the  right-most column).  This implies that the Rank 2 pattern mainly captures condition-related effects, with $C_2$ and $C_3$ inducing stronger time-increasing low-frequency activities compared to $C_1$.


\section{Discussion}  \label{sec:discussion}
In this paper, we propose a Bayesian framework for modeling multilevel two-way functional data, motivated by analyzing  time-frequency EEG responses collected under multiple experimental conditions. 
The proposed model incorporates fixed and random effects, capturing structured variability at the within-subject, between-subject, and between-condition levels, while accounting for the influence of covariates.  Through a covariate-dependent CP decomposition, the proposed model enables estimation of the marginal contributions from time and frequency domains and efficiently captures interpretable latent structures (i.e.,  ``base'' patterns) shared across subjects and conditions.  The Bayesian formulation, combined with the proposed posterior sampling algorithm and sparsity-inducing prior, provides an efficient approach for uncertainty quantification and CP rank selection. We demonstrated the advantages of our method in a simulation study, showing improved performance over existing approaches. In the EEG  alcoholism case study, we found stronger gamma-band activity in alcoholic subjects relative to controls, along with differing levels of time-increasing low-frequency activity across stimulus conditions.

While this work focuses on time-frequency responses from a single region of interest, the proposed framework can be extended to multi-region data by generalizing the CP decomposition and tensor-product basis to higher-order tensors.   Such an extension would allow for a more comprehensive characterization of spatial-temporal brain dynamics in the time-frequency domain, as well as providing a foundation for investigations into time- and frequency-resolved functional connectivity. Future work will also extend the framework to accommodate longitudinal EEG data collected over multiple visits.  


\begin{funding}
The authors were supported by the National Institute of Health (NIH Grant No. 5 R01 MH099003). 
\end{funding}

\begin{supplement}
\stitle{Supplementary Material for ``Bayesian Mixed-Effects Models for Multilevel Two-way Functional Data: Applications to EEG Experiments''} \sdescription{This file includes the identifiability proof, posterior derivations, additional results from the simulation studies, and code with usage instructions for \textbf{BMEF-1} and \textbf{BMEF-2}. It is available at \url{https://github.com/xmengju/BMEF}.}
\end{supplement}


\bibliographystyle{imsart-nameyear} 
\bibliography{references_abbreviated}       





\end{document}